\documentclass[twocolumn,nofootinbib,superscriptaddress,reprint]{revtex4-1}

\usepackage{amsmath}    
\usepackage{graphicx}   
\usepackage{verbatim}   
\usepackage{color}      
\usepackage{subfigure}  
\usepackage{hyperref}   
\usepackage{amssymb}
\usepackage{graphicx,fancybox}
\usepackage{feynmp}
\usepackage{ifpdf}
\ifpdf
\fi
\makeatletter
\def\endfmffile{%
  \fmfcmd{\p@rcent\space the end.^^J%
          end.^^J%
          endinput;}%
  \if@fmfio
    \immediate\closeout\@outfmf
  \fi
  \IfFileExists{\thefmffile.mp}{\immediate\write18{mpost \thefmffile}}{}
  \let\thefmffile\relax
}
\makeatother
\usepackage{verbatim}
\usepackage{appendix}
\usepackage{float}
\newcommand{\bsig}{\mbox{\boldmath $\sigma$}}

\newcommand{\bGam}{\mbox{\boldmath $\Gamma$}}
\newcommand{\bg}{\mbox{\boldmath $g$}}

\newcommand{\bcalf}{\mbox{\boldmath ${\cal F}$}}

\newcommand{\bff}{\mbox{\boldmath $f$}}

\newcommand{\bfT}{\mbox{\boldmath $T$}}

\newcommand{\bv}{\mbox{\boldmath $v$}}

\newcommand{\bq}{{\bf q}}

\newcommand{\bk}{{\bf k}}

\newcommand{\bp}{{\bfp}}

\newcommand{\bark}{{\bar{k}}}
\newcommand{\barq}{\bar{q}}

\newcommand{\la}{\langle}
\newcommand{\ra}{\rangle}

\newcommand{\pslash}{\not\hspace{-0.7mm}p}

\newcommand{\rsl}{\not\hspace{-0.7mm}r}

\newcommand{\qslash}{\not\hspace{-0.7mm}q}

\newcommand{\ben}{\begin{displaymath}}
\newcommand{\een}{\end{displaymath}}
\newcommand{\be}{\begin{equation}}
\newcommand{\ee}{\end{equation}}
\newcommand{\bea}{\begin{array}}
\newcommand{\eea}{\end{eqnarray}}
\newcommand{\eqn}[1]{\label{#1}}
\newcommand{\eq}[1]{Eq.~(\ref{#1})}
\newcommand{\eqs}[1]{Eqs.~(\ref{#1})}
\newcommand{\fign}[1]{\label{#1}}
\newcommand{\fig}[1]{Fig.~\ref{#1}}

\newcommand{\bfp}{{\bf p}}

\newcommand{\e}{\epsilon}

\newcommand{\bs}{\begin{split}}
\newcommand{\spl}{\end{split}}

\newcommand{\munu}{^{\mu\nu}}

\newcommand{\half}{\frac{1}{2}}

\newcommand{\br}{{\bf r}}

\newcommand{\calT}{{\cal T}}

\newcommand{\bhk}{\hat{{\bf k}}}
\newcommand{\bhq}{\hat{{\bf q}}}
\newcommand{\bhe}{\hat{{\bf e}}}

\newcommand{\psl}{\pslash}

\newcommand{\ksl}{\not\hspace{-0.7mm}k}
\newcommand{\qsl}{\qslash}

\newcommand{\barf}{\bar{f}}

\newcommand{\omkpq}{\omega_{k+q}}

\newcommand{\omk}{\omega_k}

\newcommand{\muui}{M_{a m_t'm_t}^{\bar{u}u}(\lambda)}
\newcommand{\muuj}{M_{a m_t'm_t}^{\bar{u}u}(\lambda)}

\begin{document}

\title{Electromagnetic currents of the pion-nucleon system}


\author{T. Skawronski}
\affiliation{School of Chemical and Physical Sciences, Flinders University, South Australia}
\author{B. Blankleider}
\affiliation{School of Chemical and Physical Sciences, Flinders University, South Australia}
\author{A. N. Kvinikhidze}
\affiliation{Razmadze Mathematical Institute, Republic of Georgia}



\begin{abstract}

Pion photoproduction amplitudes are calculated from a set of equations that have been derived by coupling an external photon to all places in a dressed pion-nucleon vertex.  The calculation is consistent with gauge invariance, charge conservation, unitarity and covariance.  To provide input to the photoproduction amplitude, a photon-nucleon vertex is calculated from a set of equations derived by complete attachment of photons to a dressed nucleon propagator.  We check the accuracy of this vertex by extracting its nucleon electromagnetic form factors.



\end{abstract}

\maketitle


\section{Introduction}
Photons have long been used to probe the structure of nucleons and to obtain information about their excited states \cite{Klempt:2009pi,Tiator:2011pw}.  This is done by using photons to initiate various nucleon reactions, and then analysing the results with theoretical models.  
Pion photoproduction is an example of such a reaction.  In the literature, amplitudes for this process are calculated by using integral equations to sum up an infinite number of $\gamma N\rightarrow\pi N$ Feynman diagrams, or their non-relativistic equivalents \cite{Nozawa:1989pu, Lee:1991pp, Surya:1995ur, Pascalutsa:2004pk, Matsuyama:2006rp, JuliaDiaz:2007fa}.  The equations used in these models are similar to the Bethe-Salpeter or Lippmann-Schwinger equations for $\pi N\rightarrow\pi N$, but with the incoming pion lines replaced with photons.  Although in many cases they provide a good description of the $\gamma N\rightarrow\pi N$ data, these equations are not without difficulties.  One problem is that they do not conserve electromagnetic (EM) current in a way that is theoretically correct.  

Amplitudes that have gauge invariance, and hence conserve EM current, satisfy the Ward-Takahashi identity (WTI) \cite{Takahashi:1957xn,Bentz:1986nq}.  This relates an $n+1$ point function ${\cal G}^\mu$ that has one external photon [an example of which is the five point function in \eq{thebigmommacurrent}]
 to an $n$ point Green function ${\cal G}$ [such as the four point function in \eq{thefourpointfunctionisgivenby}].  If ${\cal G}^\mu$ were evaluated exactly it would include the infinite number of diagrams that can be constructed by attaching an external photon line (EM current) to the diagrams comprising ${\cal G}$.  Summing all these terms is extremely difficult, however, and for numerical calculations to be practical one must resort to truncating the series for ${\cal G}^\mu$ and ${\cal G}$.  To preserve gauge invariance this should be done in such a way as to maintain the relationship between $ {\cal G}^\mu$ and ${\cal G}$ that is specified by the WTI.
 
There are many ways of constructing a ${\cal G}^\mu$ that satisfies the WTI, but to be consistent with the exact case, ${\cal G}^\mu$ should include the sum of all diagrams that can be obtained by attaching a photon to the diagrams retained in the approximated ${\cal G}$.  
By using various approximations, the photoproduction amplitudes of \cite{Nozawa:1989pu, Lee:1991pp, Surya:1995ur, Pascalutsa:2004pk, Matsuyama:2006rp, JuliaDiaz:2007fa} all satisfy the WTI, and hence achieve gauge invariance, to varying degrees.  However since the photons do not attach to the other particles in all possible ways, gauge invariance is not achieved in the correct manner.


To construct a $\gamma N\rightarrow \pi N$ amplitude that {\em does} achieve gauge invariance in the correct way, one can take any set of $N\rightarrow\pi N$ diagrams, write down terms for every possible way a photon could attach to them, and add them up.  The gauging of equations method \cite{talk, Haberzettl:1997jg, Kvinikhidze:1997wn, Kvinikhidze:1997gd, KB3, GPS} allows this to be done even if we want the photoproduction amplitude to contain an infinite number of diagrams.  It uses the fact that complete attachment of photons to all terms generated by an integral equation can be achieved by attaching to only a finite number of places in the integral equation itself.   The result is a closed expression for a gauge invariant, non-perturbative amplitude $T^\mu$ that can be solved in a straightforward way.  This amplitude also obeys Watson's theorem, and thus has unitarity.

In section \ref{photoproduction}, $T^\mu$ is calculated using a covariant model of the strong interaction as input.  
As in references \cite{Nozawa:1989pu, Lee:1991pp, Surya:1995ur, Pascalutsa:2004pk, Matsuyama:2006rp, JuliaDiaz:2007fa}, we work in the context of the traditional few body meson theory, where the structure of mesons and baryons is described by cutoff form factors.  This choice is both technically convenient, and relevant to current approaches used to study nucleon resonances \cite{Sato:2009de}.  Similar calculations have been undertaken by Haberzettl et al \cite{Haberzettl:2006bn,Huang:2011nx, Huang:2011as}, but in their work the dressed photon-nucleon vertex that appears in the nucleon pole term of $T^\mu$ (first diagram on the RHS of \fig{ovDSF}) was constructed by using its general analytical form as prescribed by Ball and Chiu \cite{Ball:1980ax,Ball:1980ay}.
We, on the other hand, have included in this vertex the complete sum of diagrams obtained by gauging a dressed nucleon propagator.  
To our knowledge, the $\gamma N\rightarrow\pi N$ calculations presented in this paper are the first in which gauge invariance is achieved through the complete attachment of photons to an infinite set of Feynman diagrams.  






\section{The gauging of equations method}\label{thegaugingofequationsmethod}\label{thegaugingequationsmethod}
We begin by outlining the gauging method and its application to the pion-nucleon system.  
Following reference \cite{KB3}, we consider the task of gauging the following four-point Green function ${\cal G}$ that describes interactions between point-like pions and nucleons:
\be\bs
\hspace{-0.4cm}(2\pi)^4&\delta^4(p_1'+p_2'-p_1-p_2){\cal G}(p_1',p_2',p_1,p_2)\\&=\int d^4y_1 d^4y_2 d^4x_1d^4x_2  e^{i(p_1'\cdot y_1+p_2'\cdot y_2-p_1\cdot x_1-p_2\cdot x_2)}\\&\times\la\la0|T[\psi(y_2)\phi(y_1)\bar{\psi}(x_2)\bar{\phi}(x_1)]|0\ra\ra\eqn{thefourpointfunctionisgivenby}
\end{split}\ee 
where $\psi$ and $\phi$ are Heisenberg fields of nucleons and pions, respectively, $|0\ra\ra$ is the physical vacuum and $T$ is the time ordering operator.    The $x_m$ and $y_m$ are the initial and final spacetime coordinates of pions ($m=1$) and nucleons ($m=2$), while the $p_m$, $p_m'$ are their initial and final state momenta.  When evaluated with Wick's theorem, ${\cal G}$ can be written as the sum of all possible $\pi N\rightarrow\pi N$ Feynman diagrams.  In turn, these can be split into the sums of 1 particle reducible (1PR) and 1 particle irreducible parts as ${\cal G}={\cal G}_\text{1PR}+G$.  

The gauged version of $G$ is needed in the derivation of $T^\mu$, and we first apply the gauging method to this sum of diagrams. 
It can be expressed in compact form by the Bethe-Salpeter (BS) equation:
\be\bs
G(p_1'&,p_2',p_1,p_2)=G_0(p_1',p_2',p_1,p_2)\\&+\int \frac{d^4r_1}{(2\pi)^4}\frac{d^4s_1}{(2\pi)^4}G_0(p_1',p_2',r_1,r_2)v(r_1,r_2,s_1,s_2)\\&\,\,\,\,\,\,\,\,\times G(s_1,s_2,p_1,p_2)\eqn{thisisasymbolicnotaion}
\end{split}\ee
where the total momentum of the pion-nucleon system is $p=p_1'+p_2'=p_1+p_2=s_1+s_2=r_1+r_2$.  In \eq{thisisasymbolicnotaion}, $G_0$ is the sum of all fully disconnected $\pi N\rightarrow \pi N$ diagrams and the potential $v$ is the sum of all amputated, connected, 2 particle irreducible $\pi N\rightarrow\pi N$ diagrams.  That $G_0$ is disconnected means it splits into two single particle propagators $g_N$ and $g_\pi$ which are the sums of all possible $N\rightarrow N$ and $\pi\rightarrow\pi$ diagrams:
\be\bs
&G_0(p_1',p_2',p_1,p_2)=(2\pi)^4\delta^4(p_1'-p_1)g_\pi(p_1)g_N(p_2)\eqn{equationtenksdf}
\end{split}\ee
No total momentum conservation delta function has been included in $G_0$ because the momenta are understood to be related in the way specified just below \eq{thisisasymbolicnotaion}.  

It is convenient to suppress the integrals and momentum labels in \eq{thisisasymbolicnotaion} and to write the BS equation in the shorthand notation,
 \be\bs
 G=G_0+G_0vG\eqn{thetopologicalstatement}
 \end{split}\ee
In this form, the equation is reduced to a topological statement about the structure of the Feynman diagrams belonging to $G$.  As such, it can be utilised directly to express the structure of the same Feynman diagrams, but with a photon (EM current) attached to all places in all of them.  Using a superscript $\mu$ to indicate quantities that have had this attachment carried out in all possible ways, it immediately follows that,
\be\bs
G^\mu=G_0^\mu+G_0^\mu vG+G_0v^\mu G+G_0vG^\mu\eqn{illustrateswhatwemeanbygaugingequations}
\end{split}\ee
The third term on the RHS of this equation, for instance, is shorthand for
\be\bs
&\int \frac{d^4r_1}{(2\pi)^4}\frac{d^4s_1}{(2\pi)^4}G_0(k_1,k_2,s_1,s_2)v^\mu(s_1,s_2,r_1,r_2)\\&\,\,\,\,\,\,\,\times G(r_1,r_2,p_1,p_2)\eqn{theexpression}
\end{split}\ee
The total momentum to the right of the attachment point is $p=p_1+p_2=r_1+r_2$ and that to the left is $p+q=s_1+s_2=k_1+k_2$.  The momentum $q$ is that transferred to the particles during the attachment.  

Equation (\ref{illustrateswhatwemeanbygaugingequations}) expresses the gauged version of $G^\mu$ in terms of an integral equation and illustrates what is meant by ``gauging an equation".
Both $G^\mu$ and $G_0^\mu$ are obtained from $G$ and $G_0$ but with a photon attached to all possible places in all diagrams contributing to them.  The gauged potential $v^\mu$ is similarly obtained from $v$, but because $v$ consists of amputated diagrams, $v^\mu$ does not include terms that can be obtained by attaching to their external legs.  Notice that the final three terms in \eq{illustrateswhatwemeanbygaugingequations} can also be expressed as $[G_0vG]^\mu$, which illustrates a rule for the gauging of products that is identical to the product rule for derivatives.  

Some simple algebra allows \eq{illustrateswhatwemeanbygaugingequations} to be formally solved, obtaining, \be\bs & \,\,\,\,\,\,\,\,\,\,\,\,\,\,\,\,G^\mu=-G[G^{-1}]^\mu G=G[G_0^{-1}G_0^\mu G_0^{-1}+v^\mu] G
\eqn{equationsixteen}
\end{split}\ee
where, in longhand, the inverse of $G_0$ is,
\be\bs
G_0^{-1}(p_1',p_2',p_1,p_2)=(2\pi)^4\delta^4(p_1'-p_1)g_\pi^{-1}(p_1)g_N^{-1}(p_2)
\end{split}\ee
The terms that make up the single particle propagators can also be generated by integral equations, allowing $G_0$ to be gauged in a similar way to $G$.  This is done in section (\ref{gaugeinvariantvertex}).


\subsection{The Ward-Takahashi identity}
To verify that the diagrams comprising $G^\mu$ come in the right combination to preserve gauge invariance, let's suppose that instead of constructing a five point function by gauging, it is found by evaluating,
\be\bs
&{\cal G}_\text{Exact}^{\mu}(k_1,k_2,p_1,p_2)\\&\,\,\,\,\,\,=\int d^4y_1 d^4y_2d^4x_1d^4x_2   e^{i(k_1\cdot y_1+k_2\cdot y_2-p_1\cdot x_1-p_2\cdot x_2)}\\&\,\,\,\,\,\,\,\,\,\,\,\,\times\la\la0|T[\psi(y_2)\phi(y_1)\bar{\psi}(x_2)\bar{\phi}(x_1)J^{\mu}(0)]|0\ra\ra\eqn{thebigmommacurrent}
\end{split}\ee
where the initial and final state momenta are related to each other the same way as in expression (\ref{theexpression}). 
The EM current $J^\mu$ corresponds to a phase transformation of the charged particle fields.  The generators of this transformation are,
\be\bs
&\lambda_N=\frac{e}{2}(1+\tau_3)\,\,\,\,\,\,\,\,\,\,\,\,\,\,\,\,\,\,\,\,\,\,\lambda_{\pi ba}=ie\e_{b3a}\eqn{transformationgenerators}
\end{split}\ee
where $\tau_3$ is the Pauli matrix for the third component of isospin, $\e_{b3a}$ is the Levi-Civita symbol, and $e=\sqrt{4\pi/137}$ is the elementary charge.  The $a$ and $b$ subscripts on $\lambda_\pi$ label the pion charges.

For the EM current to be conserved it must satisfy the condition $\partial_\mu J^\mu=0$ and this, in turn, results in ${\cal G}_\text{Exact}^\mu$ satisfying the two body WTI \cite{Bentz:1986nq}:
\be\bs
&-iq_\mu {\cal G}_\text{Exact}^{\mu}(k_1,k_2,p_1,p_2)\\&={\color{white}\left[\frac{}{}\right.}\lambda_\pi {\cal G}(k_1-q,k_2,p_1,p_2)+\lambda_N{\cal G}(k_1,k_2-q,p_1,p_2)\\&\,\,\,\,\,\,\,\,-{\cal G}(k_1,k_2,p_1+q,p_2)\lambda_\pi-{\cal G}(k_1,k_2,p_1,p_2+q)\lambda_N\eqn{thewardtakahashiidentityforthebigg}
\end{split}\ee
To simplify this expression, the pion charge labels have been left off.

The Feynman diagrams that comprise ${\cal G}_\text{Exact}^\mu$ come in two types.
The first type, ${\cal G}^\mu$, can be constructed by attaching photons to the Feynman diagrams that make up ${\cal G}$.  Diagrams of the second type can't be constructed in this way, but vanish when they are contracted with $q_\mu$ and so do not contribute to the WTI.  
The diagrams that make up ${\cal G}^\mu$ may be further sorted into two sets ${\cal G}_\text{1PR}^\mu$ and $G^\mu$ according to whether they can be obtained by attaching photons to ${\cal G}_\text{1PR}$ or $G$.  Thus ${\cal G}_\text{1PR}^\mu$ and $G^\mu$ must satisfy separate WTI's that add up to identity (\ref{thewardtakahashiidentityforthebigg}).  These are the same as the WTI for ${\cal G}^\mu$, but involve $G^\mu$, $G$ and ${\cal G}_\text{1PR}^\mu$, ${\cal G}_\text{1PR}$ instead.
It's easy to show that the $G^\mu$ of \eq{equationsixteen} satisfies the appropriate WTI provided that $v^\mu$ satisfies,
\be\bs
&-iq_\mu v_{lj}^{\mu}(k_1,k_2,p_1,p_2)\\&=\lambda_{\pi lm}v_{mj}(k_1-q,k_2,p_1,p_2)+\lambda_Nv_{lj}(k_1,k_2-q,p_1,p_2)\\&\,\,-v_{lm}(k_1,k_2,p_1+q,p_2){\lambda}_{\pi mj}-v_{lj}(k_1,k_2,p_1,p_2+q){\lambda}_N \eqn{wtiforthegaugedpotential}
\end{split}\ee
and $G_0^\mu$ obeys an identity the same as \eq{thewardtakahashiidentityforthebigg} but involving $G_0^\mu$ and $G_0$ instead.

To obtain an expression for $G_0^\mu$, we can use the product rule to write,
\be\bs
G_0^\mu(k_1,k_2,&p_1,p_2)=(2\pi)^4\delta^4(k_1-p_1)g_N^\mu(k_2,p_2)g_\pi(k_1)\\&\,\,\,\,\,\,\,+(2\pi)^4\delta^4(k_2-p_2)g_N(k_2)g_\pi^\mu(k_1,p_1)\eqn{usetheproductruletowrite}
\end{split}\ee
where it's understood that $k_1+k_2=p_1+p_2+q$. 
Deriving $G_0^\mu$ is thus a matter of gauging the single particle propagators.  This we do in the next section.
\subsection{Gauge invariant $\gamma N\rightarrow N$ vertex}\label{gaugeinvariantvertex}
In what follows it is assumed that the interaction of pions and nucleons is described by an interaction Lagrangian linear in the pion field.  We shall, however, neglect explicit dressings of pions. Consequently, $g_\pi$ is a Feynman propagator for a meson with mass $m_\pi\approx 138$ MeV:
\be\bs
g_\pi(k)=\frac{i}{k^2-m_\pi^2+i\e}
\end{split}\ee
Meanwhile, the sum of diagrams comprising the dressed nucleon propagator $g_N$ can be represented by the Dyson-Schwinger equation,
\be\bs
g_N(p)=g_{N0}(p)+g_{N0}(p)\Sigma_N(p) g_N(p) \eqn{swingingdysonequation}
\end{split}\ee
where the nucleon self energy $\Sigma_N$ is the sum of all $N\rightarrow N$ diagrams that are 1 particle irreducible.
The ``bare" nucleon propagator $g_{N0}$ is a Feynman propagator for a fermion with mass $m_{N0}$:
\be\bs
g_{N0}(p)=\frac{i}{\psl-m_{N0}+i\e}
\end{split}\ee
  We refer to $m_{N0}$ as the bare nucleon mass and it is set so that $g_N$ has a pole at $p_0=m_N\approx 939$ MeV in the nucleon rest frame.   We express the dressed nucleon propagator in terms of positive and negative energy components as
  \be\bs
  -ig_N(p)=\Lambda^+g_N^+(p_0)+\Lambda^-g_N^-(p_0)\eqn{douchebag:abrandnewequation}
  \end{split}\ee
  where $\Lambda^\pm=\half(1\pm\gamma_0)$.  
  The residue of $g_N^+(p_0)$ at $p_0=m_N$ is called the renormalisation constant $Z_2$ and this is calculated in section \ref{gaugeinvariantcoupledchannelsdescriptionof}.  

Equation (\ref{swingingdysonequation}) has the same form as the BS equation for $G$, so gauging $g_N$ gives,
\be\bs
&\,\,\,\,\,\,\,\,\,\,\,\,\,\,\,\,g_N^\mu=-g_N[g_N^{-1}]^\mu g_N=g_N\Gamma_{N}^\mu g_N\\& 
\Gamma_{N}^\mu=\Gamma_{N0}^\mu+\Sigma_N^\mu  \,\,\,\,\,\,\,\,\,\,\,\,\,\,\,\,\,\,g_{N0}^\mu=g_{N0}\Gamma_{N0}^\mu g_{N0}\eqn{itseasytoseethatgauginggigives}
\end{split}\ee
By analogy with the WT identities written above, $g_{N0}^\mu$ and $g_\pi^\mu$ should satisfy,
\be\bs
&-iq_\mu g_{N0}^\mu(p',p)={\lambda}_Ng_{N0}(p)-g_{N0}(p')\lambda_N\\&
-iq_\mu g_{\pi}^\mu(k',k)={\lambda}_\pi g_{\pi}(k)-g_{\pi}(k')\lambda_\pi\eqn{thebarepropagatorwtis}
\end{split}\ee
The gauged pion propagator may also be written as $g_\pi^\mu=g_\pi\Gamma_\pi^\mu g_\pi$ and it's easy to verify that the bare vertices \be\bs&\Gamma_{N0}^\mu=\lambda_N\gamma^\mu\\&\Gamma_\pi^\mu(k',k)=\lambda_\pi(k'^\mu+k^\mu)\eqn{thebarephotonvertices}\end{split}\ee allow $g_\pi^\mu$ and $g_{N0}^\mu$ to satisfy (\ref{thebarepropagatorwtis}).
Note that these vertices differ from those specified by Feynman rules (see, for example, Appendix A of \cite{Peskin:1995ev}) by a factor of $-i$.  Thus the quantities we derive by gauging need to be multiplied by $-i$ to make them consistent with Feynman rules.

To gauge the infinite number of diagrams that make up the self energy, we note that they can be expressed as,
\be\bs
&\Sigma_N(p)=\int \frac{d^4r}{(2\pi)^4}\frac{d^4s}{(2\pi)^4}\barf_0^a(k,p,p-r)\\&\,\,\,\,\,\,\,\,\,\times G_0(r,p-r,s,p-s)f^a(s,p-s,p)\eqn{theselfenergyinfull}
\end{split}\ee
The bare pion absorption vertex $\barf_0$ may be read directly from the Lagrangian, while the dressed pion emission vertex $f$ is given by the integral equation,
\be\bs
&f^a(k,p-k,p)=f_0^a(k,p-k,p)\\&\,\,\,\,\,\,\,+\int\frac{d^4r}{(2\pi)^4}\frac{d^4s}{(2\pi)^4}v_{ab}(k,p-k,r,p-r)\\&\,\,\,\,\,\,\,\,\times G_0(r,p-r,s,p-s) f^b(s,p-s,p)\eqn{thedressedpinnvertex}
\end{split}\ee
It is convenient to express the above two equations in the shorthand form,
\be\bs
&\Sigma_N=\barf_0G_0f=\barf_{0}Gf_0\\&
f=f_0+vG_0f
\eqn{thesigmainshorthandform}
\end{split}\ee
Applying the product rule then yields the gauged self energy:
\be\bs
\Sigma_N^\mu=\barf G_0^\mu f+\barf G_0f_0^\mu+\barf_0^\mu G_0f+\barf G_0v^\mu G_0 f  \eqn{biglongexpressionforgammamu}
\end{split}\ee
The vertex that results from using this in $\Gamma_N^\mu$ is illustrated in \fig{ovDSFx}.  The gauged bare vertices $f_0^\mu$, $\barf_0^\mu$ are model dependent and expressions for these are not given until section \ref{gaugingthestronginteractionmodel}.
We can however say that since $f_0$ is an amputated quantity, its gauged version needs to satisfy a WTI analogous to (\ref{wtiforthegaugedpotential}):
\be\bs
&-iq_\mu f_0^{\mu a}(k,p'-k,p)=\lambda_Nf_0^{a}(k,p'-k-q,p)\\&\,\,\,\,\,\,\,\,\,\,\,+\lambda_{\pi ab}f_0^b(k-q,p'-k,p)-f_0^a(k,p'-k,p')\lambda_N\eqn{barevertexwti}
\end{split}\ee
Meanwhile, since $g_{N0}^\mu$ satisfies (\ref{thebarepropagatorwtis}), one would expect $g_N^\mu$ to have the WTI, 
\be\bs
-iq_\mu g_N^\mu(p+q,p)=\lambda_N g_N(p)-g_N(p+q)\lambda_N\eqn{somethingweneedtogetitself}
\end{split}\ee
By putting the preceding two WTI's into $q_\mu \Gamma_N^\mu$, it's easy to recover \eq{somethingweneedtogetitself}.  Therefore, provided $f_0^\mu$ and $v^\mu$ are constructed such that they satisfy the appropriate WTI's, $g_N^\mu$ constitutes a gauge invariant description of $\gamma N\rightarrow N$.  It then immediately follows that the $G_0^\mu$ found by gauging satisfies a WTI the same as \eq{thewardtakahashiidentityforthebigg} but with $G_0^\mu$ and $G_0$ substituted for ${\cal G}^\mu$ and ${\cal G}$.

Since there exists data for the process described by $g_N^\mu$, this quantity can be compared with experiment.  First, though, the external legs of its constituent Feynman diagrams need to be ``amputated".  This involves removing each external $g_\pi$, and replacing each external $g_N$ with a factor of $\sqrt{Z_2}$.  We also adopt a convention whereby amplitudes are multiplied by an extra factor of $i$ and this cancels the $-i$ needed to convert gauged quantities to Feynman diagrams.  The properly normalised photon-nucleon vertex is therefore,
\be\bs
\Gamma_Z^\mu=Z_2\Gamma_N^\mu
\end{split}\ee

{ \begin{figure}[t] 
\centering
\includegraphics[width=8.8cm]{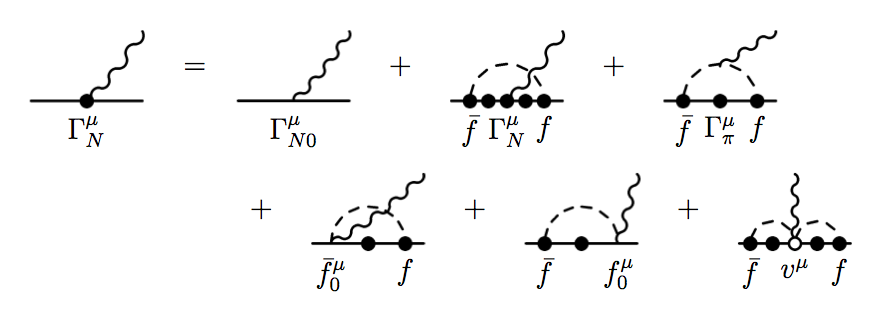}
 \caption{ The photon-nucleon vertex of \eq{itseasytoseethatgauginggigives}.  The dashed, solid, and wiggly lines denote pions, nucleons, and photons, respectively.}
\fign{ovDSFx}
\end{figure}} 

\subsection{Gauge invariant description of $\gamma N\rightarrow\pi N$}\label{andthereinliestherub}
A gauge invariant amplitude for $\gamma N\rightarrow\pi N$ may be derived by applying the gauging method to the unamputated $\pi NN$ vertex ${\cal F}=G_0fg_N=Gf_0g_N$.  The result, illustrated in \fig{ovDSF}, is given by,
\be\bs
{\cal F}^\mu&=G_0T^\mu g_N\eqn{thephotoproductionone}
\end{split}\ee
where
\begin{subequations}\begin{align}
&T^\mu=v_u^\mu+v_t^\mu+v_s^\mu+tG_0^\mu f+tG_0v^\mu G_0f\nonumber\\&\,\,\,\,\,\,\,+tG_0f_0^\mu+f_0^\mu+v^\mu G_0f\eqn{shangalang}\\&
t=v+vG_0t\\&
v_t^{\mu a}=\Gamma_\pi^\mu(k_f,k_f-q)g_\pi(k_f-q)f^a(k_f-q,p_f,p_i)
\\&
v_u^{\mu a}=\Gamma_N^\mu(p_f,p_f-q)g_N(p_f-q)f^a(k_f,p_f-q,p_i)\eqn{thecrossedborntermii}\\&
v_s^{\mu a}= f^a(k_f,p_f,p_i+q)g_N(p_i+q)\Gamma_N^\mu(p_i+q,p_i) 
\eqn{vbsparklingcii}
\end{align}\eqn{wannaspillthebloodofahippy}\end{subequations}
In writing \eq{thephotoproductionone}, use has been made of the fact that $G_0^{-1}G=1+tG_0$.  The $v^\mu$'s in equations (\ref{wannaspillthebloodofahippy}) are, like $T^\mu$, all functions of $k_f$, $p_f$, $p_i$.  The $a$ index, which labels the charge of the emitted pion, has also been dropped in many instances in order to simplify the notation.

When the inputs $v^\mu$, $f_0^\mu$, $g_N^\mu$ and $G_0^\mu$ satisfy the Ward-Takahashi identities written thus far in this section, the amplitude ${\cal F}^\mu$ is also gauge invariant.  It obeys the following WTI:
\be\bs
&-iq_\mu {\cal F}^{\mu a}(k_f,p_f,p_i)=\lambda_{\pi ab}{\cal F}^b(k_f-q,p_f,p_i)\\&\,\,\,\,\,\,\,\,\,\,\,\,\,\,\,\,\,\,\,+\lambda_N{\cal F}^a(k_f,p_f-q,p_i)-{\cal F}^a(k_f,p_f,p_i+q)\lambda_N
\end{split}\ee
The singly gauged amplitude is also consistent with two body unitarity.  To see this, we need only note that $T^\mu$ can be rearranged into a form similar to the Bethe-Salpeter and Lippmann-Schwinger equations:
\be\bs
&T^\mu=V^\mu+TG_0V^\mu\\&
V^\mu=v^\mu G_0f+f_0^\mu+\Gamma_\pi^\mu g_\pi f+f_0g_{N0}\Gamma_{N0}^\mu\\&\,\,\,\,\,\,\,\,\,\,\,+\Gamma_N^\mu g_N f+f_0g_{N0}\barf_0^\mu G_0f\\&
T=t+fg_N\barf
\end{split}\ee
In other words, $T^\mu$ has the same form as the amplitudes in \cite{Nozawa:1989pu, Lee:1991pp, Surya:1995ur, Pascalutsa:2004pk, Matsuyama:2006rp, JuliaDiaz:2007fa}.  Since they all have two body unitarity, it immediately follows that $T^\mu$ does as well.

Upon amputating the external legs from ${\cal F}^\mu$, we arrive at the following properly normalised singly gauged amplitude which incorporates gauge invariance and unitarity:
\be\bs
T_Z^{\mu a}=Z_2T^{\mu a}
\end{split}\ee

{ \begin{figure}[t] 
\centering
\includegraphics[width=7.3cm]{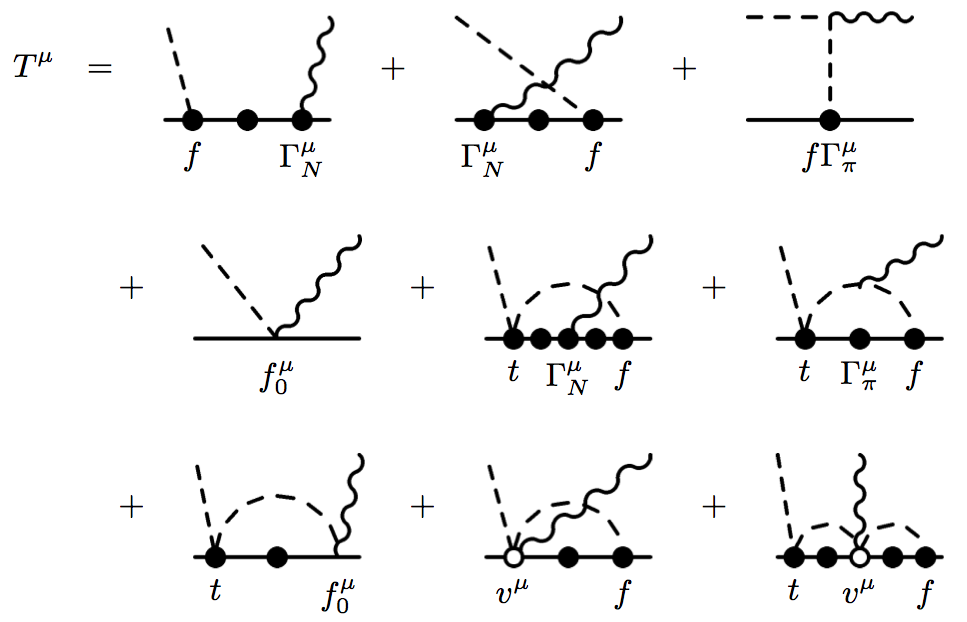}
 \caption{ The singly gauged amplitude of \eq{shangalang}.}
\fign{ovDSF}
\end{figure}}

\section{Nucleon resonances}\label{gaugeinvariantcoupledchannelsdescriptionof}
In this section, the equations for $\Gamma_Z^\mu$ and $T_Z^{\mu a}$ are extended to include contributions from excited state nucleons.
The lowest energy excited state is the $\Delta(1232)$ particle, which has spin and isospin both equal to $3/2$.  However, the lowest order $\gamma N\rightarrow\pi N$ diagram involving the $\Delta$, which is shown in \fig{youcant}, can't be obtained by gauging because the diagram that results from removing the incoming photon doesn't exist.  This is not a deficiency of the gauging method, though, because the diagram is gauge invariant on its own and can be added on to $T^\mu$ without violating the WTI.  
{ \begin{figure}[b] 
\centering
\includegraphics[width=1.8cm]{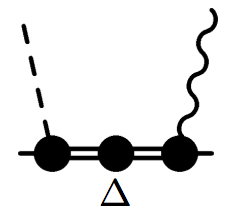}
 \caption{ Diagram needed to describe certain $\gamma N\rightarrow\pi N$ data, but which can't be obtained by gauging.}
\fign{youcant}
\end{figure}} 

Another excited state of relative importance is the $N^*(1440)$ Roper resonance.  This has the same spin and isospin as the ground state nucleon, and if the $\Delta$ particle in the intermediate state of \fig{youcant} is replaced by a Roper, the result is a class of diagrams that are not self gauge invariant, but which {\em can} be obtained by gauging.  This is done in sections \ref{andthereinliestherubii} and \ref{wevegoneoutandweveboughtstuff}.

Following the $\pi N\rightarrow\pi N$ models of \cite{PhysRevC.34.991,Gross:1992tj}, we suppose that the Roper resonance occurs in one particle states only.  For $\pi N$ scattering, this situation may be described by the properly normalised ``coupled channels" amplitude $\calT$ given by,
\be\bs
&\,\,\,\,\,\,\,\,\,\,\,\,\,\,\,\,\,\,\,\,\,\,\calT=iZ_2t+iZ_2\sum_{\alpha\beta} f_\beta g_{\beta\alpha}\barf_\alpha\\&
f_\alpha=f_{\alpha0}+t G_0f_{\alpha0}\,\,\,\,\,\,\,\,\,\,\,\,\,\,\,\,\,\,\,\,\,\,\,\,\,\barf_\alpha=\barf_{\alpha0}+\barf_{\alpha0}G_0 t\\&
\Sigma_{\alpha\beta}=\barf_{\alpha 0}G_0 f_\beta\,\,\,\,\,\,\,\,\,\,\,\,\,
g_{\alpha\beta}=g_{\alpha0}\delta_{\alpha\beta}+\sum_\rho g_{\alpha 0}\Sigma_{\alpha\rho} g_{\rho\beta}\eqn{extendedtoroperparticles}
\end{split}\ee
where the subscripts $\alpha$, $\beta$, $\rho$ can be $N$ (for a nucleon) or $R$ (for a Roper particle).  The bare Roper propagator $g_{R0}$ is the same as $g_{N0}$ but with the bare Roper mass $m_{R0}$ substituted for $m_{N0}$.  Now, Lorentz invariance tells us that the only things $\Sigma_{\alpha\beta}(p)$ could be comprised of are $\psl$ and ordinary numbers. Therefore, all elements $g_{\alpha\beta}$ of the $2\times2$ matrix $\bg$ commute and the dressed baryon propagator is,
\be\bs
\bg=\left(\begin{array}{cc}
g_R^{-1}/\Delta&
\Sigma_{NR}/\Delta\\ \Sigma_{NR}/\Delta&
g_N^{-1}/\Delta\end{array}\right)=\left(\begin{array}{cc} g_{NN} & g_{NR} \\ g_{RN} & g_{RR}\end{array}\right)
\eqn{themixedpropagatorshurryup}
\end{split}\ee
where $\Delta=g_N^{-1}g_R^{-1}-\Sigma_{NR}^2$ is the determinant of $\bg^{-1}$.   The ``no baryon mixing" Roper propagator $g_R$ is given by an expression the same as (\ref{swingingdysonequation}) but with $g_{R0}$ and $\Sigma_{RR}$ in place of $g_{N0}$ and $\Sigma_N\equiv\Sigma_{NN}$, respectively.  

 
When working in the centre of mass frame where the total momentum is $p=(p_0,{\bf 0})$, the positive energy part of $\bg$ can be isolated by writing it as,
\be\bs
&-i\bg(p)=\bg^+(p_0)\Lambda^++\bg^-(p_0)\Lambda^-
\eqn{intopositiveandnegativecomponents}
\end{split}\ee
In the coupled channels system, the elements of $\bg^+$ need to have poles at $m_N$ and the Roper mass $m_R\approx 1365-95i$ MeV.  To arrange this, the denominator
\be\bs
\Delta^+=[g_N^{-1}]^+[g_R^{-1}]^+-[\Sigma_{NR}^+]^2\eqn{positivenergypartofthedeterminant}
\end{split}\ee
must vanish when $p_0=m_N$ or $p_0=m_R$.  The quantities on the RHS of \eq{positivenergypartofthedeterminant} are defined by,
\be\bs
&ig_\beta^{-1}(p)=\Lambda^+[g_\beta^+(p_0)]^{-1}+\Lambda^-[g_\beta^-(p_0)]^{-1}\\&
i\Sigma(p)=\Sigma^+(p_0)\Lambda^++\Sigma^-(p_0)\Lambda^-
\end{split}\ee
Setting $\Delta^+(m_N)=\Delta^+(m_R)=0$ gives us a pair of simultaneous equations that can be solved to find the required bare masses and we get,
\be\bs
&m_{\alpha0}=\frac{1}{2A_\alpha}\left[-B_\alpha\pm\sqrt{B_\alpha^2-4A_\alpha C_\alpha}\right]
\eqn{thequadraticsolutionforthebaremasses}
\end{split}\ee
where
\be\bs
&A_\alpha=a_{\beta\beta}-a_{\beta\alpha}\\&
B_\alpha=(a_{\beta\alpha}-a_{\beta\beta})(a_{\alpha\beta}+a_{\alpha\alpha})-c_\alpha+c_\beta\\&
C_\alpha=a_{\alpha\alpha}a_{\alpha\beta}(a_{\beta\beta}-a_{\beta\alpha})-a_{\alpha\alpha}c_\beta+a_{\alpha\beta}c_\alpha\\&a_{\alpha\beta}=m_\beta-\Sigma_{\alpha\alpha}^+(m_\beta)\,\,\,\,\,\,\,\,\,\,\,\,\,c_\alpha=[\Sigma_{NR}^+(m_\alpha)]^2\eqn{thefirstthreelines}
\end{split}\ee
In the first three lines of \eq{thefirstthreelines}, $\beta=R$ if $\alpha=N$ and vice versa.
Meanwhile, the signs in \eq{thequadraticsolutionforthebaremasses} need to be chosen so that it reduces to $[d_R^+(m_R)]^{-1}=0$ and $[d_N^+(m_N)]^{-1}=0$ when $\Sigma_{NR}\rightarrow0$ and hence there is no mixing of the ground and excited states. 

When calculating $\Sigma^+(m_R)$, one needs to take into account that because $\Sigma^+$ is a function of $p_0=\sqrt{p^2}$, it has two possible values at each point in the complex $p^2$ plane.  To make it a single valued function, the $p^2$ plane can be replaced by a Riemann surface consisting of two sheets \cite{Pearce:1984ca,Afnan:1991kb}.  $\Sigma^+$ contributes a cut to the $p^2$ plane running just below the real axis from $(m_N+m_\pi)^2-i\e$ to $\infty-i\e$ and the two sheets can be joined along this cut so that passing through it takes us from one sheet to the other.  Then a particular $p^2$ value is on the first sheet if $0\le\arg(p^2)<2\pi$ and on the second if $2\pi\le\arg(p^2)<4\pi$.  
We must place $m_R^2$ on the second sheet when calculating $\Sigma^+(m_R)$ because an $m_R^2$ on the first sheet has $\sqrt{m_R^2}=-m_R$.  This can be done by rotating the loop integral contour such that the cut in the $p^2$ plane passes clockwise over $m_R^2$ and ends up at some angle in between the positive real axis and the negative imaginary axis.  This rotation is discussed further in section \ref{stronginteractionmodel}.

One must also ensure that the pion-nucleon interaction described by $f_\alpha$ has the correct strength.   To do this the model parameters should be adjusted such that the pion-nucleon coupling constant $g_{\pi NN}$, an expression for which is given in Appendix \ref{fixingthecouplingconstant}, has its experimental value of $13.02$.  
 
In evaluating $g_{\pi NN}$, a coupled channels renormalisation constant needs to be calculated by taking the residue of $\bg^+(p_0)$ at $p_0=m_N$.  Because $\Sigma_{NR}^+(m_N)=[g_R^{+}(m_N)]^{-1}[g_N^{+}(m_N)]^{-1}$, this residue can be factorised into the product of a row matrix and its transpose.  The ``square root" of the renormalisation constant in the coupled channels system is thus either $1\times2$ or $2\times1$:
\be\bs
\mathop{\text{Res}}_{p_0=m_N} \bg^+(p_0)&
=\left(\begin{array}{c} \sqrt{Z_N} \\ \sqrt{Z_R} \end{array}\right)\left(\begin{array}{cc} \sqrt{Z_N} & \sqrt{Z_R}\end{array}\right)\eqn{splittingthepropagatorintoarowandcolumn}
\end{split}\ee
The elements are given by,
\be\bs
&{Z_\alpha}={\frac{[g_\beta^+(m_N)]^{-1}}{\Delta^{+'}(m_N)}}\eqn{thezfactorsaretheresidues}
\end{split}\ee
where $\beta=R$ if $\alpha=N$ and vice versa. 
The prime on $\Delta^+$ in \eq{thezfactorsaretheresidues} denotes a derivative with respect to $p_0$.  
Note that when $\Sigma_{NR}\rightarrow0$ we have $[g_N^+(m_N)]^{-1}\rightarrow0$ and the usual textbook renormalisation constant is recovered:
\be\bs
&{Z_N}\rightarrow{Z_2}=\left[\left.\frac{d}{dp_0} [g_N^+(p_0)]^{-1}\right|_{p_0=m_N}\right]^{-1}\eqn{theusualrenormalisationconstant}
\end{split}\ee

 
\subsection{Gauge invariant, coupled channels description of $\gamma N\rightarrow N$}\label{andthereinliestherubii}
To obtain a gauge invariant, coupled channels $\gamma N\rightarrow N$ vertex, we repeat the process described in section \ref{gaugeinvariantvertex} but with the ordinary numbers replaced by matrices.  Doing this obtains,
\be\bs
 \bg^\mu&=-\bg[\bg^{-1}]^\mu\bg=
\bg\left(\begin{array}{cc} -[g_N^{-1}]^\mu & \Sigma_{NR}^\mu \\ \Sigma_{RN}^\mu & -[g_R^{-1}]^\mu \end{array}\right)\bg\,\,\,\,\,\eqn{thegaugedpropagatormatrix}
\end{split}\ee
The matrix between the factors of $\bg$ on the RHS will be denoted as $\bGam^\mu$.
Its elements are given by expressions the same as those in equations (\ref{itseasytoseethatgauginggigives}) and (\ref{biglongexpressionforgammamu}) but with different vertices at the edges of the pion loops:
\begin{subequations}\eqn{ljsdfjaoijosjobarbecue}
\begin{align}
&\Gamma_\alpha^\mu=-[g_\alpha^{-1}]^\mu=\Gamma_{\alpha0}^\mu+\Sigma_{\alpha\alpha}^\mu\eqn{barbecuea}\\&
\Sigma_{\beta\alpha}^\mu=\barf_{\beta0}^\mu G_0 f_\alpha+\barf_\beta G_0^\mu f_\alpha+\barf_\beta G_0 v^\mu G_0 f_\alpha+\barf_\beta G_0 f_{\alpha0}^\mu\eqn{barbecueb}
\end{align}
\end{subequations}
We have chosen the bare Roper vertex $\Gamma_{R0}^\mu$ to be the same as the bare nucleon vertex in \eq{thebarephotonvertices}.

As was the case for $g_N^\mu$, the external propagators need to be amputated from $\bg^\mu$ before it can be compared with experimental data.  That is, the external propagators must be removed and the result multiplied from the left and right by the $1\times2$ and $2\times 1$ renormalisation constants.  Thus the properly normalised coupled channels vertex is,
\be\bs
\Gamma^\mu_Z&=\left(\begin{array}{cc} \sqrt{Z_N} & \sqrt{Z_R}\end{array}\right)\left(\begin{array}{cc} -[g_N^{-1}]^\mu & \Sigma_{NR}^\mu \\ \Sigma_{RN}^\mu & -[g_R^{-1}]^\mu\end{array}\right)\left(\begin{array}{c} \sqrt{Z_N} \\ \sqrt{Z_R} \end{array}\right)
\eqn{therenormalisedgammannviaropervertex}
\end{split}\ee

\subsection{Gauge invariant, coupled channels description of $\gamma N\rightarrow\pi N$}\label{wevegoneoutandweveboughtstuff}
As the present model has Roper particles appearing only in one body states, the task of deriving a singly gauged amplitude that includes Roper contributions is a matter of gauging $\bcalf=G_0\bff\bg=G\bff_0 \bg$, where
\be\bs
\bff=\left(\begin{array}{cc}f_N&f_R\end{array}\right)\,\,\,\,\,\,\,\,\,\,\,\,\,\,\,\,\,\,\,\,\,\,\,\,\,\,\,\,\bff_0=\left(\begin{array}{cc}f_{N0}&f_{R0}\end{array}\right)
\end{split}\ee
The product rule yields an expression for $\bcalf^\mu$ that is very similar to \eq{thephotoproductionone}:
\be\bs
\bcalf^\mu=G_0\bfT^\mu \bg
\end{split}\ee
where
\be\bs
&\bfT^\mu=\bv_u^\mu+\bv_t^\mu+\bv_s^\mu+tG_0^\mu \bff+tG_0v^\mu G_0\bff\\&\,\,\,\,\,\,\,+tG_0\bff_0^\mu+\bff_0^\mu+v^\mu G_0\bff\eqn{thephotoproductionamplitude}
\end{split}\ee
and
\begin{subequations}\begin{align}
&\bv_t^{\mu a}=\Gamma_\pi^\mu(k_f,k_f-q)g_\pi(k_f-q)\bff^a(k_f-q,p_f,p_i)
\\&
\bv_u^{\mu a}=\Gamma_N^\mu(p_f,p_f-q)g_N(p_f-q)\bff^a(k_f,p_f-q,p_i)\eqn{thecrossedbornterm}\\&
\bv_s^{\mu a}= \bff^a(k_f,p_f,p_i+q)\bg(p_i+q)\bGam^\mu(p_i+q,p_i) 
\eqn{vbsparklingc}
\end{align}\eqn{vbsparkling}\end{subequations}
The gauged matrix vertex $\bff_0^{\mu a}$ is the same as $\bff_0^a$, but with $f_{N0}^{\mu a}$, $f_{R0}^{\mu a}$ substituted for $f_{N0}^a$, $f_{R0}^a$.  To allow the $g_N$ inside $G_0$ to have a nucleon pole, the bare mass inside the two body propagator $G_0$ is taken to be different to that in $\bg$.    
Thus, amputating the external legs of $\bcalf^\mu$ now involves removing the external propagators, multiplying from the right by the $2\times1$ renormalisation constant and from the left by $\sqrt{Z_2}$:
\be\bs
T_Z^{\mu a}&=\sum_\rho\sqrt{Z_2}T_\rho^{\mu a}\sqrt{Z_\rho}\eqn{thebigropermixingoverlandhastwoindianpacificcoachesonittoday}
\end{split}\ee

\section{Strong interaction model}\label{stronginteractionmodel}

To provide the necessary input to a calculation of $\Gamma_Z^\mu$ and $T_Z^{\mu a}$, the bare vertices $f_{\alpha0}$ and the potential $v$ need to specified.
Rather than developing a new model for these, we have borrowed from the covariant $\pi N\rightarrow\pi N$ description due to Gross and Surya (GS) \cite{Gross:1992tj} which is based on equations (\ref{extendedtoroperparticles}).  In common with most other models that use these and similar equations, the two particle intermediate states are simplified by replacing $G_0$ with bare, physical mass propagators multiplied by $Z_2$.  The factor of $Z_2$ causes the approximated propagator to have the same nucleon pole residue as its exact counterpart.  To further reduce technical complexities, GS chose a separable $v$ and used the spectator approach to reduce all loop integrals to three dimensions.

{ \begin{figure}[h] 
\centering
\includegraphics[width=6.4cm]{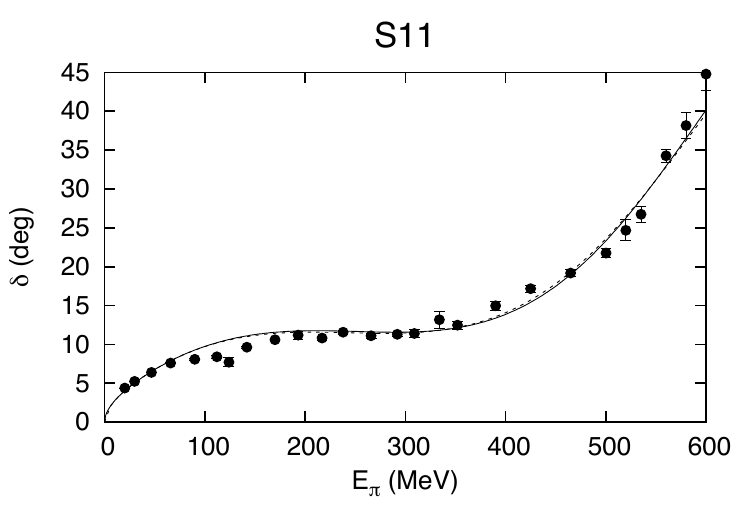}
 \includegraphics[width=6.4cm]{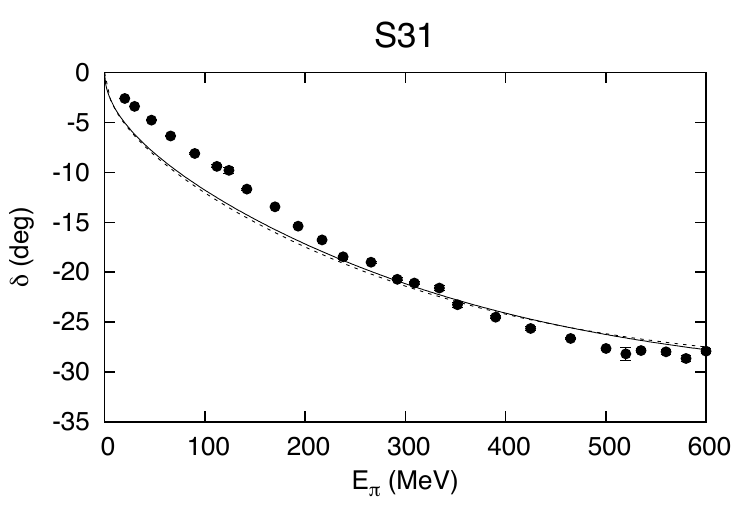}\\ 
\includegraphics[width=6.4cm]{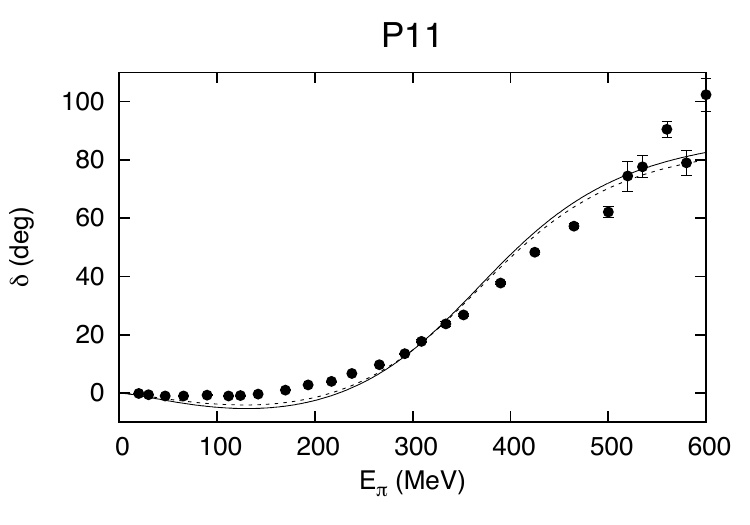}
\includegraphics[width=6.4cm]{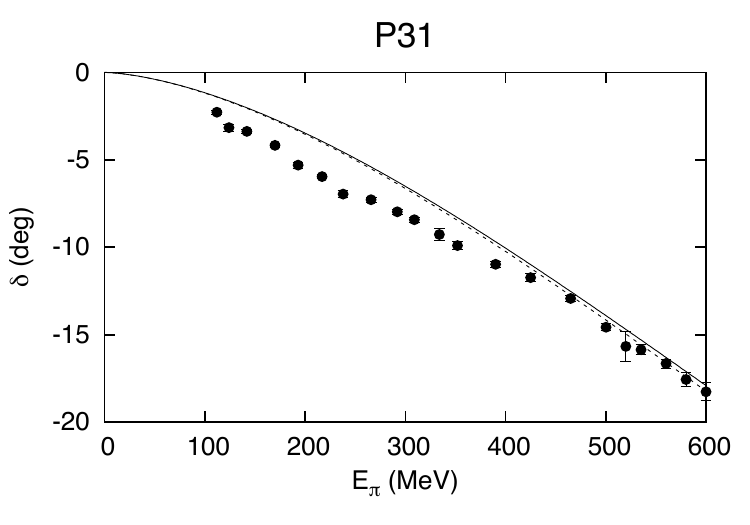}
\vspace{-0.4cm} \caption{ Phase shifts for $j=1/2$.  The Fit A results (dotted lines) are almost indistinguishable from those for Fit B (solid lines).  The data is taken from \cite{SAID}.
}
   \fign{phaseshiftsfivefourtwelvec}
\end{figure}}

GS implemented the spectator approach by placing intermediate pions on shell.  However, when we used this approximation to calculate $\Gamma_Z^\mu$, it was found to produce EM form factors that vary only slightly with the squared photon momentum.  To overcome this problem, we carried out the spectator approach by putting the intermediate nucleons on shell instead.  This amounts to making the replacement,
\be\bs
G_0&(k',r',k,r)\rightarrow(2\pi)^4\delta^4(k'-k)Z_2\frac{\pi}{E_r} i[\rsl+m_N]\\&\,\,\,\,\times\frac{\delta^+(r^2-m_N^2)}{[p_0-E_r+\omega_k-i\e][p_0-E_r-\omega_k+i\e]}\eqn{tworesiduereplacement}
\end{split}\ee
where $p=k+r=k'+r'$ and $E_{r}=\sqrt{\br^2+m_N^2}$.  
Despite this change, we retained the cutoff factors, bare pion vertices, and background potential of the GS paper.  The cutoff factors are,
\be\bs
h_\alpha(k^2)=\left(\frac{(\Lambda_\alpha^2-\tilde{m}_\alpha^2)^2}{(\Lambda_\alpha^2-\tilde{m}_\alpha^2)^2+(\tilde{m}_\alpha^2-k^2)^2}\right)^2\eqn{gandssformfactors}
\end{split}\ee
Ground state nucleons and pions have $\tilde{m}_N=m_N$, $\tilde{m}_\pi=m_\pi$ while $\tilde{m}_R$ and the cutoff masses $\Lambda_N$, $\Lambda_\pi$, $\Lambda_R$ are treated as free parameters.  The bare vertices each have a factor of $\sqrt{Z_2}$ absorbed into them, and are related to the vertices of section \ref{thegaugingequationsmethod} by $\Gamma_{\alpha0}^{\pi a}=\sqrt{Z_2}f_{\alpha0}^a$, $\bar{\Gamma}_{\alpha0}^{\pi a}=\sqrt{Z_2}\barf_{\alpha0}^a$.  They are given by,
\be\bs
&\bar{\Gamma}_{\alpha0}^{\pi a}(k,p',p)=g_{\pi N\alpha}^{(0)}\tau_a\gamma_5\left[x_\alpha-y_\alpha\ksl\right]F_\alpha(k^2,p^2,p'^2)\\&
\Gamma_{\alpha0}^{\pi a}(k,p',p)=g_{\pi N\alpha}^{(0)}\tau_a^*\left[x_\alpha-y_\alpha\ksl\right]\gamma_5 F_\alpha(k^2,p'^2,p^2)
\eqn{thebarepinnvertices}
\end{split}\ee
where
\be\bs &F_\alpha(k^2,r'^2,r^2)=h_\alpha(r^2)F_V(k^2,r'^2)\\&F_V(k^2,r^2)=h_\pi(k^2)h_N(r^2)\end{split}\ee
and $y_\alpha=(1-x_\alpha)/(2m_N)$.  The asterisk on the Pauli matrix in $\Gamma_{\alpha0}^{\pi a}$ denotes a Hermitian conjugate, while $g_{\pi N\alpha}^{(0)}$ is the bare coupling constant.  The extent to which the pion-nucleon coupling is pseudoscalar and pseudovector is governed by $x_\alpha$, which is a number between 0 and 1.

Similar to the bare vertices, the GS background potential incorporates a factor of $iZ_2$ and is related to the potential written above by $iZ_2v_{ba}={\cal V}_{ba}$.  It has the following form:
\be\bs 
{\cal V}_{ba}={\cal V}_{1/2}(P_{1/2})_{ba}+{\cal V}_{3/2}(P_{3/2})_{ba}\eqn{separablepotential}
\end{split}\ee
where
\be\bs 
&{\cal V}_t(k',r',k,r)=F_V(k'^2,r'^2)\tilde{{\cal V}}_t(p)F_V(k^2,r^2)\\&
\tilde{{\cal V}}_t(p)=C_0^t(p^2)\psl+C_1^t\eqn{thebitsofthegandspotential}
\end{split}\ee
and $p=k'+r'=k+r$.  The isospin projection operators are,
\be\bs
&(P_{1/2})_{ba}=\frac{1}{3}\tau_b^*\tau_a\\&
(P_{3/2})_{ba}=\delta_{ba}-\frac{1}{3}\tau_b^*\tau_a\eqn{isospinprojection}
\end{split}\ee
while the $C$ factors are written down in Appendix \ref{thebackgroundpotential}.

To set the parameters, $\pi N\rightarrow\pi N$ phase shifts were calculated from the renormalised amplitude $\calT_{ba}(k_f,p_f,k_i,p_i)$ using the formulae given in Appendix \ref{pionnucleonscatteringappendix}.  
The best set of parameters we could find when using replacement (\ref{tworesiduereplacement}) are listed as Fit A in Table \ref{awholebunchofuselessnumbers} and produce the phase shifts shown in \fig{phaseshiftsfivefourtwelvec}.  The horizontal axes of these graphs show the pion laboratory energies $E_\pi$, which are related to the total momentum $p=k_f+p_f=k_i+p_i$ by,
\be\bs
E_\pi=\frac{p^2-m_\pi^2-m_N^2}{2m_N}-m_\pi
\end{split}\ee

 \begin{table}[b]
 \begin{center}
 \begin{tabular}{lcc}
 \hline\hline \,\,\,\,\,\,\,\,\,\,\,\,\,\,\, & 
 \,\,\,\,A \,\,\,\,\,\,\,\,& \,\,\,\,\,\,\,\,\,\,\,\,\,\,\, B\,\,\,\,\,\,\,\,\,\,\,\,\,\,\,\,\, \\ \hline
 $\Lambda_N$ &
1432.99924
 &
1358.45006
 \\
 $\Lambda_R$ &
1948.45836
 &
1943.97350
 \\
 $\Lambda_\pi$ &
 582.51508
 &
 582.61099
 \\
 $x_N$ &
     0.17775
 &
   0.06548
 \\
 $x_R$ &
    0.79282
  &
   0.87960

 \\
 $g_{\pi NN}^{(0)}$ &
  10.99164
  &
  13.44624
 \\
 $g_{\pi NR}^{(0)}$ &
   12.13224
 &
  10.70639
 \\
 $\tilde{m}_R$ &
1449.48265
 &
1455.25030
 \\
  \hline\hline
 \end{tabular}
 \end{center}
 \vspace{-0.5cm}
   \caption{ Parameters that were adjusted to fit the $j=1/2$ phase shift data. The cutoff masses are in units of MeV, as is $\tilde{m}_R$.  The other parameters are dimensionless.  }
 \label{awholebunchofuselessnumbers}
 \end{table}

 \begin{table}[t]
 \begin{center}
 \begin{tabular}{lcc}
 \hline\hline \,\,\,\,\,\,\,\,\,\,\,\,\,\,\, & 
 \,\,\,\,A \,\,\,\,\,\,\,\,& \,\,\,\,\,\,\,\,\,\,\,\,\,\,\,\,\,\,\,\,\,\,\,\, B\,\,\,\,\,\,\,\,\,\,\,\,\,\,\,\,\,\,\,\,\,\,\,\,\,\,\,\, \\ \hline
 $\sqrt{Z_N}$ &
    1.09671
 &
   0.91862

 \\
 $\sqrt{Z_R}$ &

   0.04759
 
 &
   0.04069
 
 \\
 $\sqrt{Z_2}$ &
     1.04774

 &
   0.96018
 
 \\
 $m_{N0}$ &

1032.45740

 &
1047.36929
 
 \\
 $m_{R0}$ &

1967.19681
 &
1832.62105
 
 \\
 
 $M_R$ & $1362.9-96.8i$ & $1362.7-96.7i$
 \\

 $g_{\pi NN}$ & 
   12.79023
 
 &
  12.97247\\  \hline\hline
 \end{tabular}
 \end{center}
   \caption{ Parameters that are determined by the fits.  The masses $m_{N0}$, $m_{R0}$, $M_R$ are in units of MeV and the other parameters are dimensionless.}
 \label{awholebunchofuselessnumbersii}
 \end{table}

To facilitate calculation of the bare Roper mass, $m_R^2$ has been placed on the second sheet of the Riemann surface discussed in section \ref{gaugeinvariantcoupledchannelsdescriptionof}.
In the centre of mass frame where $p=(p_0,{\bf 0})$, the factor $[p_0-\omega_k-E_r+i\e]^{-1}$ on the RHS of replacement (\ref{tworesiduereplacement}) causes a cut in the $p^2$ plane running from $p^2=(m_N+m_\pi)^2-i\e$ to $p^2=\infty-i\e$ when $|\br|$ is integrated from $0$ to $\infty$.  This cut can be moved across $m_R^2$ by rotating the $|\br|$ integral using the replacement $|\br|\rightarrow|\br|e^{-i\phi}$, with $\phi\gtrapprox 12.5$ degrees.  The squared Roper mass then ends up on the second sheet of the $p^2$ Riemann surface, as required.

We have restricted $m_{R0}$ to being real in order to preserve unitarity.  This condition made giving $\bg^+$ a pole exactly at the Roper mass rather difficult, and the Fit A parameters cause it to have a pole at $M_R=1362.9-96.8i$ MeV, which is slightly different to $m_R$.

Another issue with Fit A is the fact that $Z_2$, which is interpreted as being a probability, is greater than 1 (see Table \ref{awholebunchofuselessnumbersii}).  In exact field theory, the only singularities that $g_N^+(p_0)$ contributes to the $p_0$ plane are a pair of cuts running from $\pm(m_N+m_\pi-i\e)$ to $\pm(\infty-i\e)$ and it can be shown that this results in $Z_2$ being restricted to the range $0\le Z_2\le 1$.  Introducing cutoff factors such as $h_\alpha$, however, gives $g_N^+(p_0)$ extra poles and this allows $Z_2$ to move outside the unitary bound.
The spectator approach also modifies the structure: when replacement (\ref{tworesiduereplacement}) is used inside $g_N^+(p_0)$, the cuts move so that they run from $p_0=0+i\e$ to $p_0=m_N-m_\pi+i\e$ and from $p_0=m_N+m_\pi-i\e$ to $p_0=\infty-i\e$ instead.  We found it impossible to obtain both a decent fit to the phase shifts and a $Z_2$ in the correct range when replacement (\ref{tworesiduereplacement}) was used.

It turns out, however, that if the spectator approach is implemented by putting both nucleons and anti-pions on shell, the correct cut structure can be preserved.  Thus, we also tried using the two-term spectator approach,
\be\bs
&G_0(k',r',k,r)\rightarrow (2\pi)^4\delta^4(k'-k)Z_2\frac{\pi i[\rsl+m_N]}{p_0-E_r+\omega_k}\\&\,\,\,\,\,\,\,\,\,\times\left[\frac{\delta^+(r^2-m_N^2)}{E_r[p_0-\omega_k-E_r+i\e]}+\frac{\delta^-(k^2-m_\pi^2)}{\omega_k[p_0+\omega_k+E_r-i\e]}\right]
\eqn{tworesiduereplacementii}
\end{split}\ee
where $\omega_k=\sqrt{\bk^2+m_\pi^2}$.  The first and second terms in the large square brackets put nucleons and anti-pions on shell and, together, contribute cuts to the $p_0$ plane that are the same as those in full four dimensional field theory.  The denominator outside the square brackets doesn't contribute any cut structure because when it passes through zero the square bracketed terms cancel.  

A set of parameters corresponding to replacement (\ref{tworesiduereplacementii}) is labelled as Fit B in Tables (\ref{awholebunchofuselessnumbers}) and (\ref{awholebunchofuselessnumbersii}).  The phase shifts for this fit are almost indistinguishable from the fit A results, and $Z_2$ falls in the correct range despite the poles contributed by the cutoff factors.

\section{Gauging the strong interaction model}\label{gaugingthestronginteractionmodel}
In this section, expressions for the gauged bare $\pi NN$ vertices, gauged potential, and gauged intermediate propagators are presented.  They are needed as input to the pion-photon vertex and pion photoproduction amplitude given in section \ref{gaugeinvariantcoupledchannelsdescriptionof}.

\subsection{Gauged bare $\pi NN$ vertex}\label{gaugedbarepinnvertex}
To gauge the bare $\pi NN$ vertices, we use the method of minimal substitution.   
The application of this procedure to $\pi NN$ vertices with cutoff factors is discussed at length in the PhD thesis of van Antwerpen \cite{antwerpenthesis} (an abridged version of which was published in \cite{vanAntwerpen:1994vh}).  By taking a combination of his results for pseudoscalar and pseudovector coupling, we can immediately write,


\be\bs
f_{\alpha 0}^{\mu i}&(k',r',r)=\Pi_{k'}^{\mu ij}[f_{\alpha0}^j(k',r',r)-f_{\alpha0}^j(k'-q,r',r)]\\&+ [f_{\alpha0}^i(k',r',r+q)-f_{\alpha0}^i(k',r',r)]\Pi_r^\mu\\&+\Pi_{r'}^\mu[f_{\alpha0}^i(k',r',r)-f_{\alpha0}^i(k',r'-q,r)]\\&-\gamma_5Z_2^{-1/2} F_V([k'-q]^2,r'^2)h_\alpha(r^2)g_{\pi N\alpha}^{(0)}y_\alpha\\&\,\,\,\,\,\,\,\times\left[i\gamma^\mu \lambda_{\pi ij}+\Pi_{k'}^{\mu ij}\qslash\right]\tau_j^*\eqn{gaugedvertexforemissionbody}
\end{split}\ee
where
\be\bs
&\Pi_{k'}^{\mu ij}=i\lambda_{\pi ij}\frac{2k'^\mu-q^\mu}{(k'-q)^2-k'^2}\\&
\Pi_r^\mu=i\lambda_N\frac{2r^\mu+q^\mu}{r^2-(r+q)^2}\\&\Pi_{r'}^\mu=i\lambda_N\frac{2r'^\mu-q^\mu}{(r'-q)^2-r'^2}\eqn{thegaugedbarevertex}
\end{split}\ee
It's easy to verify that this gauged vertex satisfies identity (\ref{barevertexwti}).
The $f_{\alpha0}^i(k',r',r)$ terms on the RHS of \eq{gaugedvertexforemissionbody} don't contribute to the WTI, but play an important role in ensuring that $f_{\alpha0}^\mu$ does not contain any poles.  Whenever the denominators in the $\Pi$ factors pass through zero, these terms cause the numerators to vanish and give a derivative.

Similarly, the gauged pion absorption vertex is given by,
\be\bs
\barf_{\alpha0}^{\mu i}&(k,r',r)=\left[{\barf^j_{\alpha0}(k+q,r',r)-\barf^j_{\alpha0}(k,r',r)}\right]\Pi_k^{\mu ji}\\&+\left[{\barf^i_{\alpha0}(k,r',r+q)-\barf^i_{\alpha0}(k,r',r)}\right]\Pi_r^\mu\\&+\Pi_{r'}^\mu\left[{\barf^i_{\alpha0}(k,r',r)-\barf^i_{\alpha0}(k,r'-q,r)}\right]\\&- \tau_jZ_2^{-1/2}g_{\pi N\alpha}^{(0)} F_V([k+q]^2,r^2)h_\alpha(r'^2)y_\alpha\\&\,\,\,\,\,\,\,\times\left[i\gamma^\mu \lambda_{\pi ji}+\Pi_k^{\mu ji}\qsl\right]\gamma_5  \eqn{gaugedvertexforabsorptionbody}
\end{split}\ee
where
\be\bs
\Pi_k^{\mu ji}=i\frac{2k^\mu+q^\mu}{k^2-(k+q)^2}\lambda_{\pi ji}\eqn{ttheextracapitalpi}
\end{split}\ee

\subsection{Gauged potential}\label{thepotential}
To gauge the background potential $v$, we begin by using the product rule on the first of \eqs{thebitsofthegandspotential}:
\be\bs
&v^{\mu}_{ji}(k',r'+q,k,r)=F_L^{\mu jit}\tilde{v}_t(p)F_R+F_{Lp}\tilde{v}_t^{\mu ji}(p+q,p) F_R\\&\,\,\,\,\,\,\,\,\,+F_{Lp}\tilde{v}_t(p+q)F_R^{\mu jit}\eqn{thegaugedpotential}
\end{split}\ee
where $p=k+r=k'+r'$ and $\tilde{v}_t=-iZ_2^{-1}\tilde{\cal V}_t$.  $F_{Lp}$ and $F_R$ are products of cutoff factors with the momentum dependence,
\be\bs
&F_{Lp}=F_V(k'^2,[r'+q]^2)\\&F_R=F_V(k^2,r^2)
\end{split}\ee 
To gauge the functions of momentum we again borrow from van Antwerpen's results and use,
\be\bs
&F_L^{\mu jit}=\left[(F_L-F_{Lk'})\Pi_{k'}^{\mu jn}+(F_L-F_{Lr'})\Pi_{r'}^\mu\delta_{jn}\right](P_t)_{ni}\\&\,\,\,\,\,\,\,\,\,\,+(F_{Lp}-F_L)\Pi_p^{\mu jit}\\&
F_R^{\mu jit}=(P_t)_{jn}\left[(F_{Rk}-F_{Rp})\Pi_k^{\mu ni}+(F_{Rr}-F_{Rp})\Pi_r^\mu\delta_{ni}\right]\\&\,\,\,\,\,\,\,\,\,\,+(F_{Rp}-F_R)\Pi_p^{\mu jit}\\&
\tilde{v}_t^{\mu ji}(p+q,p)=[\tilde{v}_t(p+q)-\tilde{v}_t(p)]\Pi_p^{\mu jit}\\&\,\,\,\,\,\,\,\,\,\,\,+iZ_2^{-1}\lambda_T^{jit}C_0^t([p+q]^2)\left[\gamma^\mu-\frac{\qsl(2p^\mu+q^\mu)}{(p+q)^2-p^2}\right]\eqn{thegaugedreducedpotentialsfoundbymakingthestretch}
\end{split}\ee
where
\be\bs
&F_{L}=F_V(k'^2,[r'+q]^2)\,\,\,\,\,\,\,\,\,\,\,\,\,\,\,\,\,\,\,\,\,\,\,\,\,\,\,\,F_{Rp}=F_V(k^2,r^2)\\&
F_{Lk'}=F_V([k'-q]^2,[r'+q]^2)\,\,\,\,\,\,\,\,\,F_{Rk}=F_V([k+q]^2,r^2)\\&
F_{Lr'}=F_V(k'^2,r'^2)\,\,\,\,\,\,\,\,\,\,\,\,\,\,\,\,\,\,\,\,\,\,\,\,\,\,\,\,\,\,\,\,\,\,\,\,F_{Rr}=F_V(k^2,[r+q]^2)
\end{split}\ee
and
\be\bs
\Pi_p^{\mu jit}=\frac{2p^\mu+q^\mu}{p^2-(p+q)^2}\lambda_T^{jit}\eqn{theextraextrapi}
\end{split}\ee
Since $p$ is the total momentum of the pion-nucleon system it has been associated with the total ``charge":\be \lambda_T^{jit}=i(\lambda_N\delta_{jn}+\lambda_{\pi jn})(P_t)_{ni}=i(P_t)_{jn}({\lambda}_N\delta_{ni}+{\lambda}_{\pi ni})\eqn{chargeconservation}\ee  
Putting the pieces together, a gauged potential that satisfies WTI (\ref{wtiforthegaugedpotential}) is obtained.
The $C$ functions in this $v^\mu$ depend only on $p$ and $q$ and do not participate in any integrals, resulting in $v^\mu$ retaining the separability of the original potential.

\subsection{Gauging on shell particles}\label{placinggaugedpionsonshell}
When the spectator approach is implemented by putting intermediate nucleons on shell, there are two possible ways of applying it to diagrams that have two internal nucleon lines within a single loop, such as the fifth diagram on the RHS of figure \ref{ovDSF}.  In order to satisfy the WTI, Gross and Riska \cite{GR} approximated diagrams such as this by replacing them with two terms.  In one term, the particle to the left of the photon vertex was put on shell, and in the other the particle to the right of the photon vertex received the same treatment.  
However, because this prescription leads to the non-conservation of charge, we will use a modified version suggested by \cite{Kvinikhidze:1997wn}:

\be\bs
G_0^\mu(k',r',k,r)\rightarrow G_N^\mu(k',r',k,r)+G_\pi^\mu(k',r',k,r)\eqn{similartothoseabove}
\end{split}\ee
where we now have $k'+r'=k+r+q=p+q$ and
\begin{widetext}
\begin{subequations}\begin{align}
&G_N^\mu(k',r',k,r)=-Z_2(2\pi)^4\delta^4(k'-k)[\rsl'+m_N]\Gamma_{N0}^\mu(r',r) [\rsl+m_N]\nonumber\\&\times\left\{\frac{\pi}{E_{r'}}\frac{\delta^+(r'^2-m_N^2)}{[E_r+q_0-E_{r'}](q_0-E_{r'}-E_r)\{p_0+q_0-E_{r'}-\omega_k+i\e\}[p_0+q_0-E_{r'}+\omega_k-iz\e]}\right.\nonumber\\&+\frac{\pi}{E_r}\frac{\delta^+(r^2-m_N^2)}{[E_r+q_0-E_{r'}](E_r+q_0+E_{r'})\{p_0-E_r-\omega_k+i\e\}[p_0-E_r+\omega_k-iz\e]}\nonumber\\&\left.+\frac{\pi}{\omega_k}\frac{\delta^-(k^2-m_\pi^2)(1-z)}{[p_0+\omega_k-E_r]\{p_0+\omega_k+E_r-i\e\}[p_0+q_0+\omega_k-E_{r'}]\{p_0+q_0+\omega_k+E_{r'}-i\e\}}\right\}\eqn{gaugingonshellparticlesi}\\&\nonumber\\&
\nonumber G_\pi^\mu(k',r',k,r)=-Z_2(2\pi)^4\delta^4(r'-r)[\rsl+m_N]\Gamma_\pi^\mu(k',k)\\&\nonumber \times\left\{\frac{\pi}{\omega_{k'}}\frac{\delta^-(k'^2-m_\pi^2)(1-z)}{[q_0+\omega_{k'}-\omega_k](q_0+\omega_{k'}+\omega_k)[p_0+q_0+\omega_{k'}-E_r]\{p_0+q_0+\omega_{k'}+E_r-i\e\}}\right.\\&\nonumber +\frac{\pi}{\omega_k}\frac{\delta^-(k^2-m_\pi^2)(1-z)}{(q_0-\omega_k-\omega_{k'})[q_0-\omega_k+\omega_{k'}][p_0+\omega_k-E_r]\{p_0+\omega_k+E_r-i\e\}}\\&+\left.\frac{\pi}{E_r}\frac{\delta^+(r^2-m_N^2)}{[p_0+q_0-E_r-\omega_{k'}+i\e][p_0+q_0-E_r+\omega_{k'}-iz\e]\{p_0-E_r-\omega_k+i\e\}[p_0-E_r+\omega_k-iz\e]}\right\}\eqn{gaugingonshellparticlesii}
\end{align}\eqn{lkdsjflkas}\end{subequations}
\end{widetext}
If prescription (\ref{tworesiduereplacement}) is used to implement the spectator approach in $G_0$, $z=1$ should be chosen; $z=0$ corresponds to prescription (\ref{tworesiduereplacementii}).  

In $G_N^\mu$ the delta function demands $k'=k$ and hence $r=p-k$, $r'=p+q-k$.  Meanwhile, $G_\pi^\mu$ has $r'=r$ and hence $r=p-k$, $k'=k+q$.
With that in mind, it's easy to see that $G_N^\mu$ and $G_\pi^\mu$ remain regular when the square bracketed denominators without $i\e$ terms go to zero.  For instance, when $q_0=\omk-\omkpq$, we have $ \omkpq+\omk+q_0=2\omk$, $q_0-\omk-\omkpq=-2\omkpq$ and $p_0+q_0+\omkpq=p_0+\omk$.  
The first two terms contained in the large curly brackets of \eq{gaugingonshellparticlesii} therefore cancel out and result in a derivative rather than a singularity.

Note that the denominators in the round brackets can't pass through zero when $q^2\le0$ (this is the only case that we will be considering). 
When $-|\bq|\le q_0\le0$ it's obviously impossible to have $\omkpq+\omega_k-q_0=0$.  If we also have a case of $|\bq|<|\bk|$ then \be-\omkpq-\omega_k<-|\bk|<-|\bq|\le q_0\ee and it's not possible for $\omega_k+\omega_{k+q}+q_0$ to pass through zero either.  Now consider the situation of $|\bq|\ge |\bk|$.  In the extreme case where $\bq$ is parallel to $\bk$ we have,
\be\bs -\omkpq-\omega_k&=-\sqrt{(|\bk|+|\bq|)^2+m_\pi^2}-\sqrt{\bk^2+m_\pi^2}\\&<-|\bq|\le q_0
\end{split}\ee
In the other extreme case where $\bk$ is anti-parallel to $\bq$ we have
\be\bs
-\omkpq-\omega_k&=-\sqrt{(|\bk|-|\bq|)^2+m_\pi^2}-\sqrt{\bk^2+m_\pi^2}\\&<-\left|\frac{}{}|\bk|-|\bq|\frac{}{}\right|-|\bk|\\&\le q_0
\end{split}\ee
Thus the zeroes of all round bracketed denominators are excluded.  A proof that is nearly identical to the one above shows that the zeros of $\pm(\omkpq+\omega_k)-q_0$ are inaccessible for $0\le q_0\le |\bq|$ as well.  


{ \begin{figure*}[ht] 
\centering
\includegraphics[width=7.0cm]{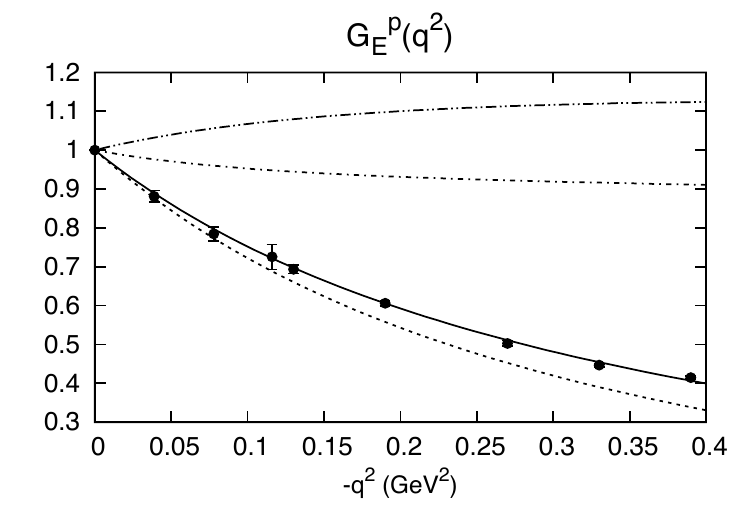}
 \includegraphics[width=7.0cm]{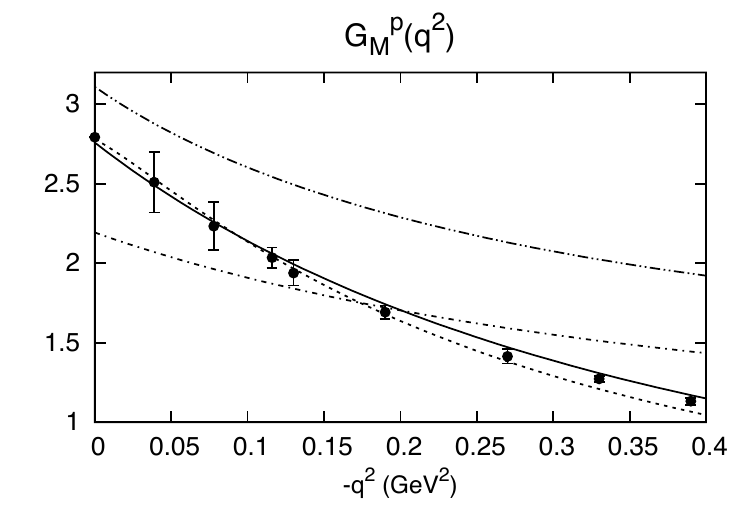}\\ 
\includegraphics[width=7.0cm]{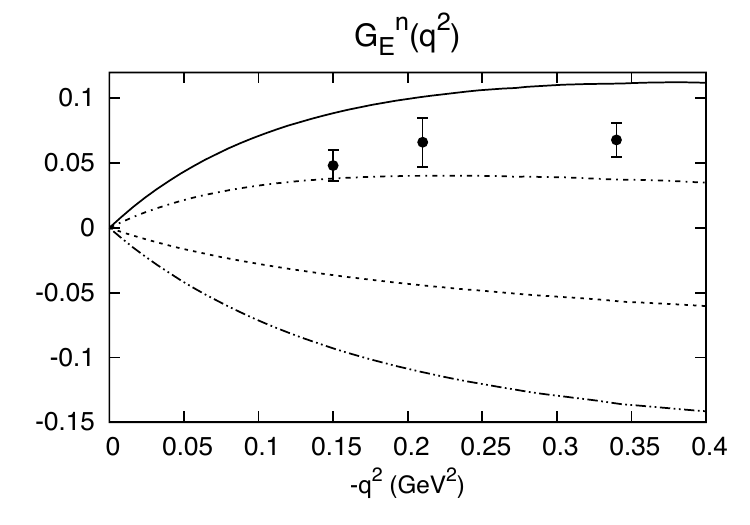}
\includegraphics[width=7.0cm]{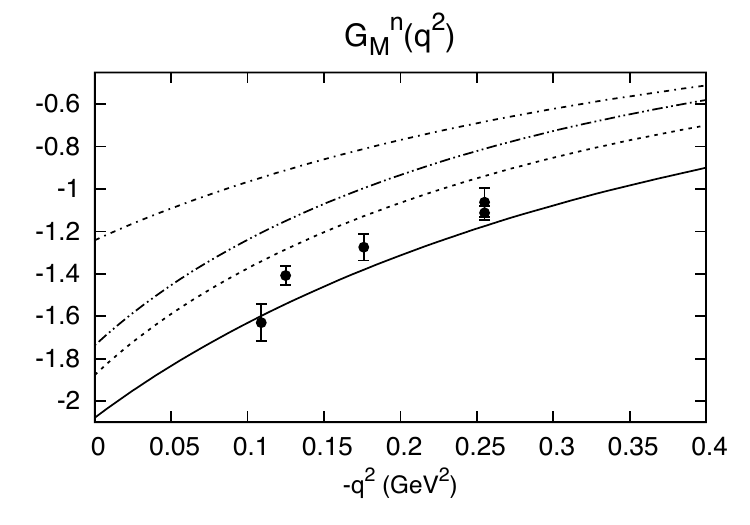}
  \caption{ Nucleon electromagnetic form factors.  The meaning of the different line types is explained in the text. }
   \fign{ffeightfourtwelvec}
\end{figure*}}

\section{Nucleon electromagnetic form factors} \label{thedressedgammannvertex}
We now have the ingredients to calculate the photon-nucleon vertex in \eq{therenormalisedgammannviaropervertex}.  To test its accuracy, nucleon electromagnetic form factors have been extracted.  This is done by using
Lorentz invariance and the Gordon identity to express the on shell version of $\Gamma_Z^\mu$ as,
\be\bs
&\bar{u}(\bp')\Gamma^\mu_Z(p',p)u(\bp)\\&=e\bar{u}(\bp')\left(F_1(q^2)\gamma^\mu+\frac{i\sigma\munu q_\nu}{2m_N}F_2(q^2)\right)u(\bp)\end{split}\ee
where it is assumed that $p^2=p'^2=m_N^2$.  The Dirac matrix coefficients $F_1$, $F_2$ are the electromagnetic form factors and are $2\times 2$ matrices in isospin space.   They can be decomposed into proton and neutron components:
\be\bs
F_{1,2}=\frac{1+\tau_3}{2}F_{1,2}^p+\frac{1-\tau_3}{2}F_{1,2}^n
\end{split}\ee
It is conventional to report the form factors in the combinations,
\be\bs
&G_E(q^2)=F_1(q^2)+\frac{q^2}{4m_N^2}F_2(q^2)\\&
G_M(q^2)=F_1(q^2)+F_2(q^2)
\end{split}\ee
which are the Sachs form factors.
Those obtained from the Fit A parameters and prescription (\ref{tworesiduereplacement}) are shown in \fig{ffeightfourtwelvec}, denoted by the dot-dot-dashed lines.  The $G_E^p$ and $G_M^p$ data is taken from \cite{PhysRevD.4.45}, the $G_E^n$ data from \cite{Herberg:1999ud, PhysRevLett.82.4988} and the $G_M^n$ data from \cite{PhysRevC.48.R5, PhysRevLett.75.21}.  That we have obtained $G_E^p(0)=1$, $G_E^n(0)=0$ is a strong indication that the Ward identity is satisfied and that gauge invariance is being maintained.  The curve for $G_M^n$ also predicts the data with reasonable accuracy.  However, the $G_E$ curves have slopes with signs opposite to what they should be, while that for $G_M^p$ has a magnitude that is too large.  

The Fit B form factors are denoted by the dot-dashed lines in \fig{ffeightfourtwelvec}.   The slopes of these curves all have the correct signs, but they are not steep enough and the magnitudes of $G_M^{n,p}(0)$ are not very good either.  

Overall, it would appear that our model, as it stands, is incapable of describing the EM form factors for $q^2<0$.  There are, on the other hand, a lot of successful models in the literature.   One early theory postulated that photons, after turning into quark-antiquark pairs, couple to nucleons as vector mesons. The vector meson propagators have a dipole form, and when enough exchange diagrams are included in these Vector Meson Dominance models, they are able to fit the form factor data very well without including any pion loops at all \cite{Iachello:1972nu, PhysRevC.69.068201}.   However, these models are essentially phenomenological, and impart little information about the underlying nucleon structure.  
Other early calculations were based on dispersion theory \cite{Nyman197097, Hare:1972zt}.  
More recently, nucleon form factor calculations have used lattice QCD \cite{Gockeler:2003ay, Alexandrou:2011db}, cloudy bag models \cite{Miller:2002ig}, quark models \cite{Bloch:1999ke,Eichmann:2011vu} and chiral perturbation theory \cite{Kubis:2000zd, Fuchs:2003ir, Schindler:2005ke}.  
The latter calculations involve pion loop diagrams, but are based on quite complicated Lagrangians.  Accordingly, the bare $\gamma NN$ vertices have a lot of physics in them.  In contrast, we have simply used $\frac{e}{2}(1+\tau_3)\gamma^\mu$ for this vertex.   

Another possible source of error is that, having removed the dressing from the 2 particle intermediate states, all $\gamma NN$ vertices appearing inside our pion loops are bare ones.  This has resulted in $\Gamma_N^\mu$ being the solution of an ordinary equation rather than an integral one.    
However, since $\Gamma_Z^\mu$ still contains an infinite number of diagrams we will assume an inadequate choice of bare $\gamma NN$ vertex is the main reason our calculations give such poor results.  Accordingly, we replaced all instances of $\Gamma_{N 0}^\mu$ with,
\be\bs
\Gamma_{N0}^\mu\rightarrow\tilde{F}_1(q^2)\gamma^\mu+\tilde{F}_2(q^2)\frac{i}{2m_N}\sigma\munu q_\nu+\tilde{F}_3(q^2)\qsl q^\mu\eqn{wtsatsifyinggoodffgiving}
\end{split}\ee
The first two coefficients are chosen to have the dipole form,
\be\bs
&\tilde{F}_1(q^2)=e\frac{1}{(1-b_1q^2)^2}\frac{1+\tau_3}{2}\\&
\tilde{F}_2(q^2)=e\frac{a_{2p}}{(1-b_{2p}q^2)^2}\frac{1+\tau_3}{2}+e\frac{a_{2n}}{(1-b_{2n}q^2)^2}\frac{1-\tau_3}{2}
\end{split}\ee
The remaining coefficient in \eq{wtsatsifyinggoodffgiving}, which is included so that the new $\Gamma_{N0}^\mu$ satisfies the WTI, is,
\be
\tilde{F}_3(q^2)=\frac{1}{q^2}\left[e\frac{1+\tau_3}{2}-\tilde{F}_1(q^2)\right]
\ee
Since $\tilde{F}_1(0)=\frac{e}{2}(1+\tau_3)$, $\tilde{F}_3(q^2)$ in the limit $q^2\rightarrow0$ is a finite quantity involving the derivative of $\tilde{F}_1$ with respect to $q^2$.  With this new bare vertex, the calculation is able to achieve a good description of the data, as shown by the dashed (Fit A) and solid (Fit B) lines in \fig{ffeightfourtwelvec}.  The parameters $b_1$, $b_{2\alpha}$ and $a_{2\alpha}$ are given in table \ref{moreuselessnumbers}.

 \begin{table}[h]
 \begin{center}
 \begin{tabular}{ccc}
 \hline\hline &
 A & B  \\ \hline
 $b_1$ &
     2.000
 &
     1.408
 \\
 $b_{2p}$ &
    10.000
 &
     1.408
 \\
 $b_{2n}$ &
      1.111
&
     1.408
 \\
 $a_{2p}$ &
      -0.260
 &
     0.715
 \\
 $a_{2n}$ &
      -0.100
 &
    -0.978
 \\ \hline\hline
 \end{tabular}
 \end{center}
 \caption{  Parameters of the bare $\gamma NN$ vertex in \eq{wtsatsifyinggoodffgiving}.  The $b$ parameters have units of $\text{GeV}^{-2}$, while the $a$'s are dimensionless.}\label{moreuselessnumbers}
 \end{table}

\section{Multipole amplitudes}
\label{photoproduction}

We now come to calculating the $\gamma N\rightarrow\pi N$ amplitude of \eq{thebigropermixingoverlandhastwoindianpacificcoachesonittoday}.  
  All the parameters have already been set either by the $\pi N$ phase shifts or by the EM form factors and there is no further adjustment.   The results presented in this section are therefore all predictions of the data, rather than fits to it.

It is convenient to work in the centre of mass of the incoming photon and nucleon, and for the motion of the final state pion and nucleon to be in the $xz$ plane.  This corresponds to the kinematics $T_Z^{\mu a}(k_f,p-k_f,p_i)$, where
\be\bs
&k_f=(\omega_{\bark},\bark\sin\theta,0,\bark\cos\theta)\,\,\,\,\,\,\,\,\,\,\,\,\,\,\,\,
p_i=(E,0,0,-\barq) \\& q=(\barq,0,0,\barq)\,\,\,\,\,\,\,\,\,\,\,\,\,\,\,\, \,\,\,\,\,\,\,\,\,\,\,\,\,\,\,\,\,\,\,\,\,\,\,\,\,\,\,\,\,\,\,\,E=\sqrt{\barq^2+m_N^2}\\&
\barq=\frac{p_0^2-m_N^2}{2p_0}\,\,\,\,\,\,\,\,\,\,\,\,\,\,\,\,\,\,\,\,\,\,\,\,\,\,\,\,\,\,\,\,\,\,\,\,\,\,\,\,\,\,\,\,\,\,\,\,
p_0^2=2E_\gamma m_N+m_N^2
\eqn{howpeoplechoosethemomentumforphotproduction}
\end{split}\ee
The momenta of the incoming nucleon, outgoing pion, and the photon are denoted by $p_i$, $k_f$ and $q$, respectively.  They are all taken to be on shell.  The total momentum of the $\pi N\gamma$ system is $p=p_i+q$ and since we require $p_{0}>m_N+m_\pi$ to produce a pion the energy of the incoming photon $E_\gamma$ must be chosen to be greater than $m_\pi+m_\pi^2/(2m_N)\approx 148 \text{ MeV}$.  The on shell relative momentum is given by,
\be\bs
\bark=\sqrt{\frac{[p_0^2-(m_N+m_\pi)^2][p_0^2-(m_N-m_\pi)^2]}{4p_0^2}}\eqn{therelativemomentumonshell}
\end{split}\ee


In calculating $T_Z^{\mu a}$, the dressed propagator and photon vertex in the crossed Born term $\bv_u^{\mu a}$ have been replaced by their bare equivalents.  That is to say we have made the replacement,
\be\bs
\bv_u^{\mu a}(k',r',r)\rightarrow&\Gamma_{N0}^\mu\frac{i}{\rsl'-\qsl-m_N+i\e}\bff^a(k',r'-q,r)
\end{split}\ee
The resulting $T_Z^{\mu a}$ is the same as the one obtained by attaching photons in all possible ways to $\tilde{g}_{N0}g_\pi\bff\bg$, where $\tilde{g}_{N0}$ is the same as $g_{N0}$ but for a particle of mass $m_N$.
We tried calculating the $\bv_u^\mu$ of \eq{thecrossedbornterm} by boosting the dressed propagator and vertices to moving frames, but did not obtain very good results.   This is understandable, though, because the parameters have been set for the specific case of the dressed quantities being in stationary frames.

 { \begin{figure}[t] 
\centering
\includegraphics[width=4.3cm]{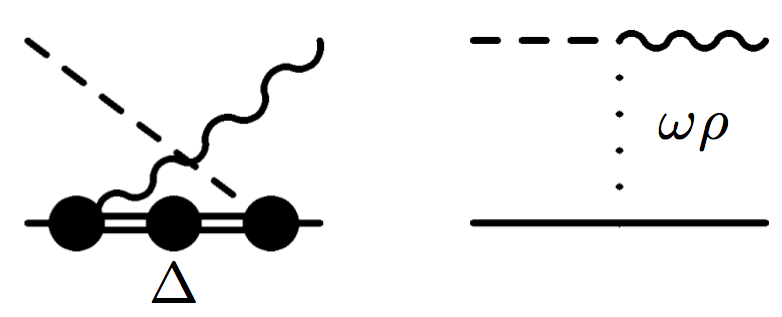}
 \caption{ Terms that contribute to $\gamma N\rightarrow\pi N$ but which are not included in our model.}
\fign{toolazy}
\end{figure}} 
 
We have also left out the $\Delta$ resonance and heavy meson exchange diagrams of figures (\ref{youcant}) and (\ref{toolazy}), even though they are needed to describe certain photoproduction data.
The $\Delta$ diagrams, for instance, have a large impact on a number of multipole amplitudes, particularly those for which the total isospin $t$ or total angular momentum $j$ are equal to $3/2$.  However, the exact size of their contribution is in part determined by the $\Delta$ electromagnetic form factors and because these are not well known, they are usually taken to be free parameters (as in, for example, \cite{Nozawa:1989pu, Pascalutsa:2004pk, Surya:1995ur}).  As we are primarily concerned with studying the ability of $T_Z^{\mu a}$ to predict, rather than simply describe, the data, we have dropped the $\Delta$ diagrams and concentrated on the amplitudes for which $j=1/2$, $t=1/2$.  The $\Delta$ is not expected to contribute much (if at all) to these channels.
 
Likewise, the meson exchange diagrams are neglected because they can also be expected to make a fairly small contribution to the $j=1/2$, $t=1/2$ amplitudes (compare the dotted and dash-dotted curves in figures 3 and 4 of reference \cite{Pascalutsa:2004pk} for an example of this).

{ \begin{figure}[b] 
\centering
\includegraphics[width=6.4cm]{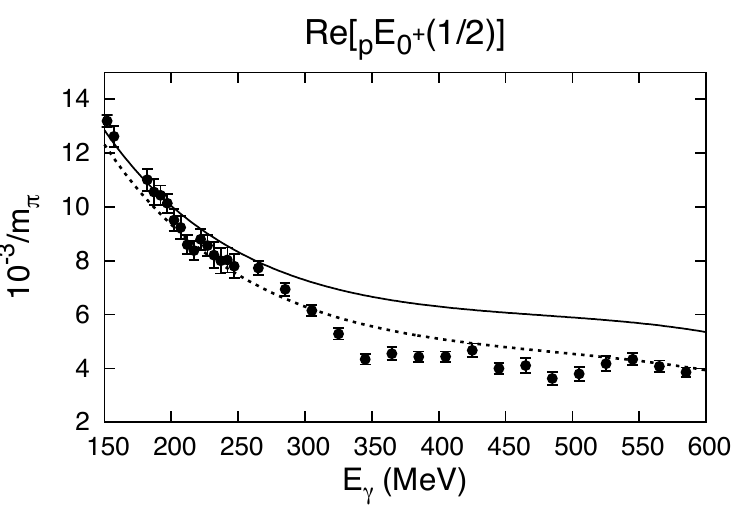}
 \includegraphics[width=6.4cm]{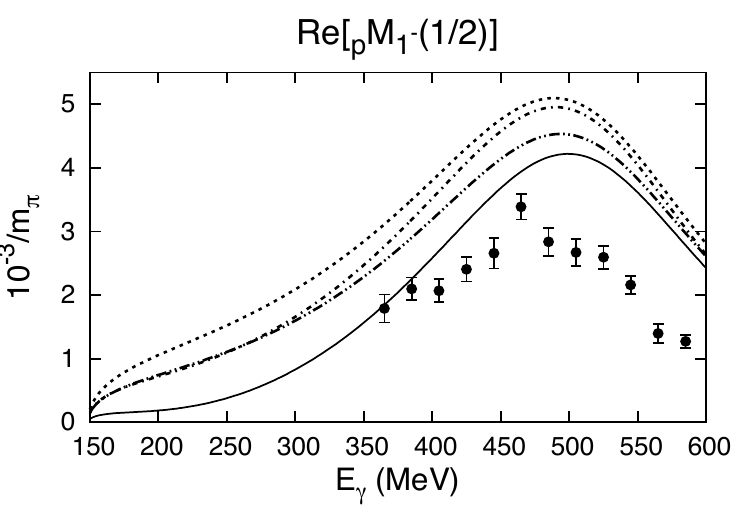}\\ 
\includegraphics[width=6.4cm]{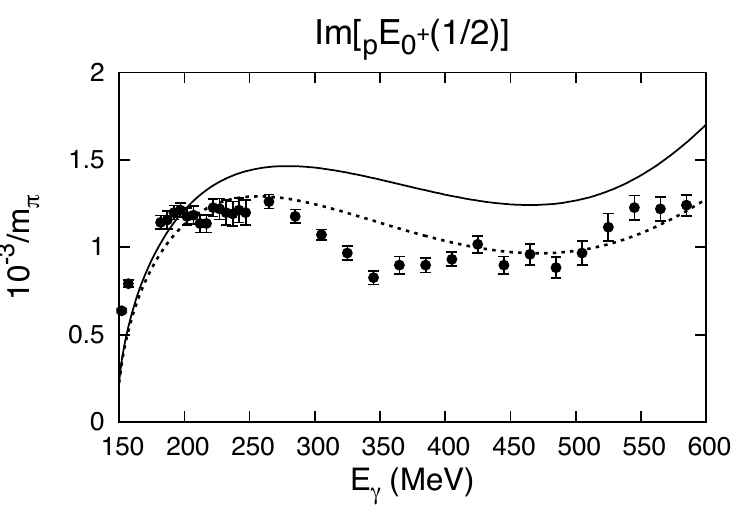}
 \includegraphics[width=6.4cm]{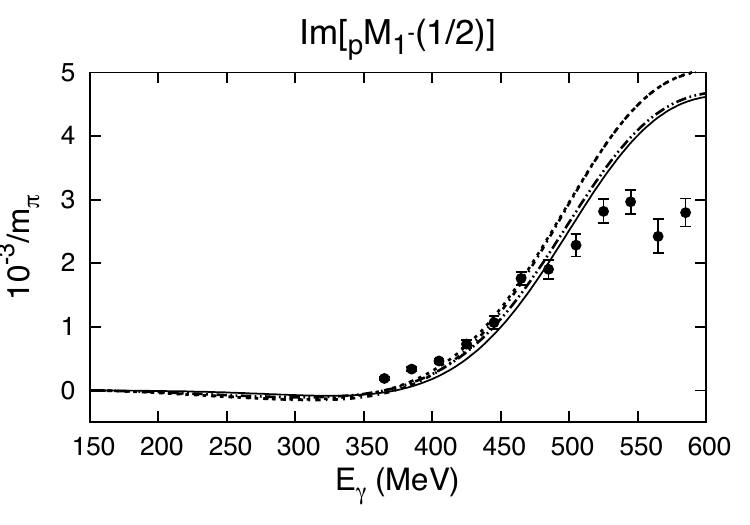}\\ 
 \caption{ Proton multipole amplitudes for $j=1/2$, $t=1/2$.  The various line styles denote the same thing as in \fig{ffeightfourtwelvec}, while the data is taken from \cite{SAID}.}
 \fign{multipolesi}
\end{figure}}

The multipole amplitudes were extracted from $T_Z^{\mu a}$ using the formulae written in Appendix \ref{pionphotoproductionappendix}.  These amplitudes are labelled with the notation $M_{l^\pm}(t)$ and $E_{l^\pm}(t)$, where $l$ is the orbital angular momentum of the emitted pion and the $\pm$ sign indicates the total angular momentum $j=l\pm\half$.  $M$ denotes magnetic multipole amplitudes and $E$ the electric ones; using \eq{drechselandtiatorsstuff} these are further decomposed into proton and neutron components which are labelled by subscript $n$'s and $p$'s.

The $j=1/2$, $t=1/2$ proton multipole amplitudes are shown in \fig{multipolesi}, where the different line styles denote the same scenarios as in \fig{ffeightfourtwelvec}.  The neutron amplitudes, shown in \fig{multipolesii}, are of similar quality.   When $\Gamma_{N0}^\mu=\half e(1+\tau_3)\gamma^\mu$ is used, the curves for Fit A are mostly quite close to the data, although those for $M_{1^-}(1/2)$ have magnitudes that are a little too large.  Parametrising the photon-nucleon vertices according to \eq{wtsatsifyinggoodffgiving} makes no appreciable difference to the $E_{0^+}(1/2)$ amplitudes,  but causes the $M_{1^-}(1/2)$ ones to become less accurate in the case of Fit A.  A possible explanation for this is that since the parametrisation has been set up to correct the on shell version of $\bGam^\mu$, the $\gamma NN$ vertices appearing in diagrams other than the uncrossed Born term $\bv_s^{\mu a}$ do not have the right on shell values (the $\gamma NN$ vertices in diagrams other than $\bv_s^{\mu a}$ are just $\Gamma_{N0}^\mu$).  It is also possible that, in addition to making the on shell form factors more accurate, the parametrisation has the side effect of making the off shell vertices less so.  

The $M_{1^-}(1/2)$ curves for Fit B, on the other hand, show an improvement when the bare photon vertex is parametrised.  Using $\half e(1+\tau_3)\gamma^\mu$ for $\Gamma_{N0}^\mu$ produces curves that are not too bad, although their magnitudes are a little too large at energies greater than about 300 MeV.  Parametrising $\Gamma_{N0}^\mu$, however, moves the $M_{1^-}$ curves downward and makes them more accurate.   



{ \begin{figure}[t] 
\centering
 \includegraphics[width=6.4cm]{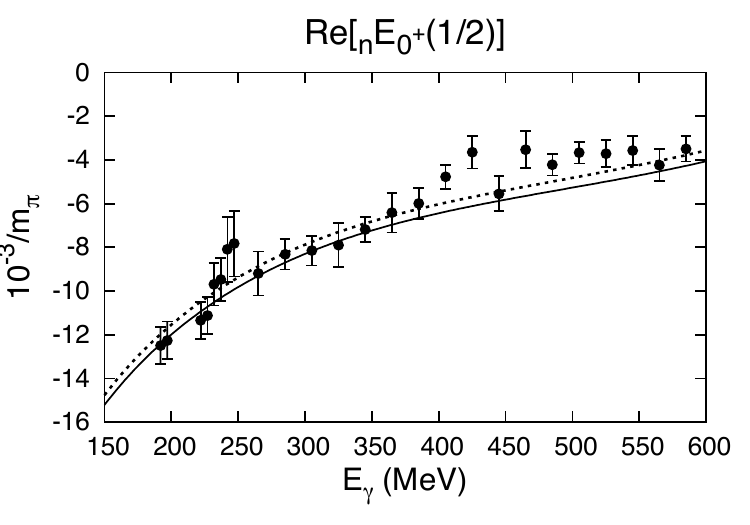}
 \includegraphics[width=6.4cm]{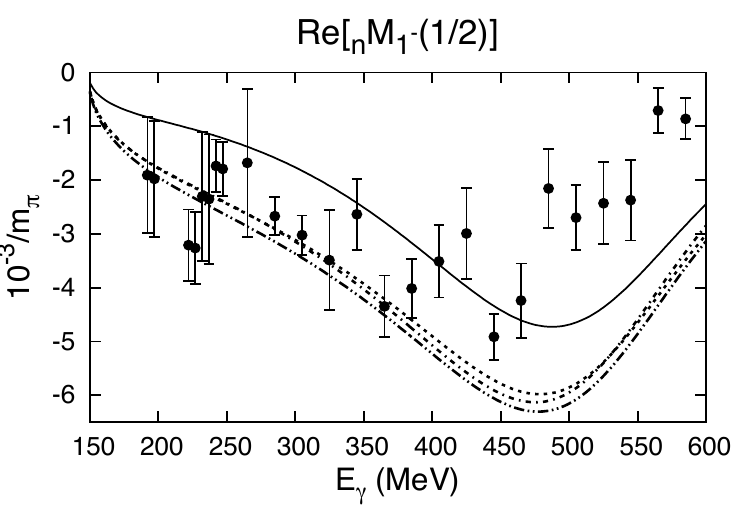}\\ 
\includegraphics[width=6.4cm]{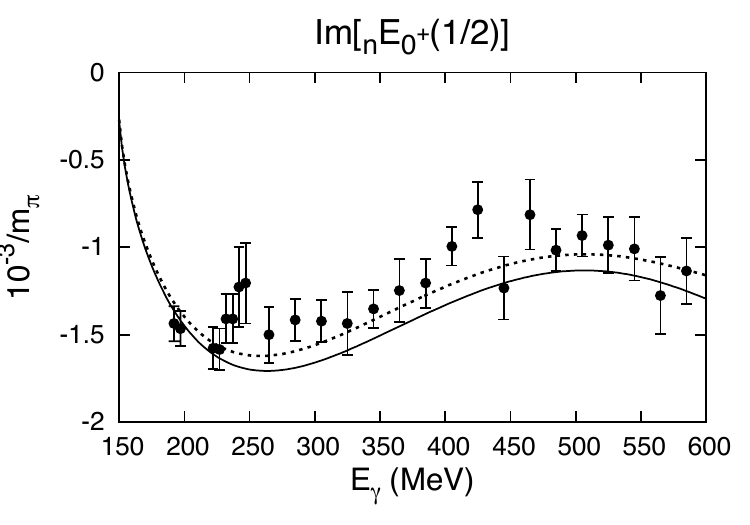}
 \includegraphics[width=6.4cm]{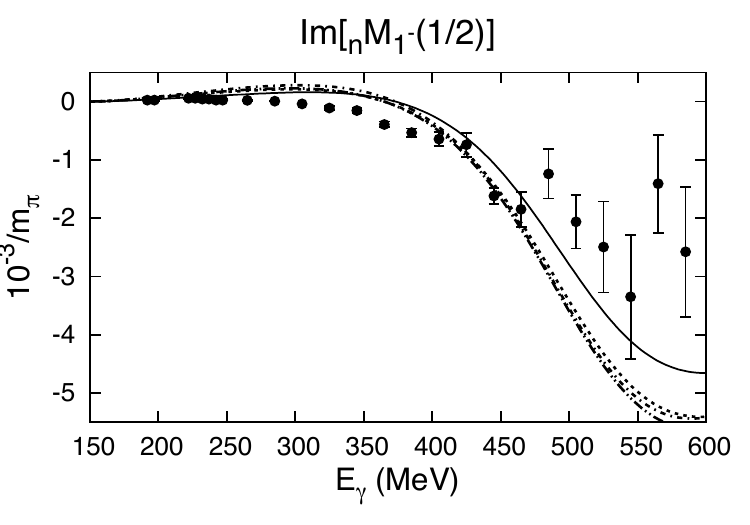}\\ 
 \caption{ Neutron multipole amplitudes for $j=1/2$, $t=1/2$.  The various line styles denote the same thing as in \fig{ffeightfourtwelvec}.}
 \fign{multipolesii}
\end{figure}}

\section{Conclusion}

In this paper we have calculated a $\gamma N\rightarrow\pi N$ amplitude that has both unitarity and gauge invariance.  For the first time, gauge invariance has been achieved through the complete attachment of photons to an infinite set of $N\rightarrow\pi N$ diagrams.  

 In these covariant initial investigations, we have chosen to reduce loop integrals to three dimensions using the covariant spectator approach.
 The results obtained so far are encouraging: there have been no insurmountable technical difficulties and the $\gamma N\rightarrow\pi N$ amplitudes predict photoproduction data for $j=1/2$, $t=1/2$ fairly well (these are the channels to which only diagrams involving pions and nucleons are expected to make significant contributions).  The model can be easily extended to describe amplitudes for $j>1/2$, $t>1/2$ through the addition of extra nucleon resonances and heavy meson exchange diagrams.
 
Meanwhile, the dressed photon-baryon vertex $\bGam^\mu$ that is an input to the $\gamma N\rightarrow\pi N$ calculation gives a reasonable prediction of the nucleon electromagnetic form factors for on shell ($q^2=0$) photons.  To get a good description of the form factors for $q^2<0$, we found it necessary to parametrise the bare photon-nucleon vertex.  Happily, this parametrisation also improved the quality of the Fit B $M_{1^-}(1/2)$ multipole amplitudes.

The numerical calculations we have presented in this paper successfully illustrate how the gauging of equations method can be used in practice and it is hoped that the results will be useful for future developments.


\begin{appendix}

\section{The $\pi N\rightarrow\pi N$ potential}

\label{thebackgroundpotential}
Although the $\pi N\rightarrow\pi N$ potential we have used is the same as that in \cite{Gross:1992tj}, it is written down here in order to standardise the notation.  
The $C$'s that appear in \eq{thebitsofthegandspotential} are given by the expressions,
\be\bs
&C_1^{1/2}=-[C_{1,N}+C_{1,R}]+C_{1,\sigma\rho}\\&
C_1^{3/2}=2[C_{1,N}+C_{1,R}]+C_{1,\sigma\rho}\\&
C_0^{1/2}=-[C_{0,N}+C_{0,R}]+4[C_{0,\sigma\rho}+C_{0,\rho}]\\&
C_0^{3/2}=2[C_{0,N}+C_{0,R}]-2[C_{0,\sigma\rho}+C_{0,\rho}]
\end{split}\ee
where
\be\bs
&
C_{0,N}(p^2)=Cg_{\pi NN}^{(0)^2}h_N^2(u)\left[\frac{1}{p^2+m_N^2-m_Nm_\pi-2m_\pi^2}\right.\\&\,\,\,\,\,\,\,\,\,\left.-\frac{(1-x_N)^2}{4m_N^2}\frac{p^2+m_\pi^2-m_N^2}{2p^2}\right]\\&
C_{0,\rho\sigma}(p^2)=-C\frac{g_{\pi NN}^{(0)^2}}{m_N}h_N^2(u)\frac{(1-x_N)^2}{4m_N}\frac{p^2+m_\pi^2-m_N^2}{2p^2}\\&
C_{0,R}(p^2)=g_{\pi NR}^{(0)2}\frac{h_{R}^2(u)}{\sqrt{p^2}}m_\pi\left[\frac{1}{{\tilde{m}_R}^2-u}-\frac{(1-x_R)^2}{(m_N+\tilde{m}_R)^2}\right]\\&
C_{0,\rho}(p^2)=-C_\rho\frac{g_{\pi NN}^{(0)^2}}{4m_N^2}h_N^2(u)\frac{p^2+m_\pi^2-m_N^2}{2p^2} \eqn{thenonprogczerosandcones}
\end{split}\ee
\be\bs
&C_{1,N}=Cg_{\pi NN}^{(0)^2}h_N^2(u)\frac{x_N^2-1}{2m_N}\\&
C_{1,\rho\sigma}=-C\frac{g_{\pi NN}^{(0)^2}}{m_N}h_N^2(u)x_N^2\\&
C_{1,R}=g_{\pi NR}^{(0)^2}h_{R}^2(u)\left[\frac{\tilde{m}_R-m_N}{{\tilde{m}_R}^2-u}+\frac{x_R^2-1}{m_N+\tilde{m}_R}\right]\\&
\end{split}\ee
and $u=(m_N-m_\pi)^2$.
Most of the parameters contained in these expressions also appear in the bare vertices and are listed in table \ref{awholebunchofuselessnumbers}.   The exceptions are $C$, $C_\rho$, which are given in table \ref{beltana}.  
 \begin{table}[b]
 \begin{center}
 \begin{tabular}{ccc}
 \hline \hline & 
 A   & B  \\ \hline
 $C$ &
      1.42929
 &
   1.00364
 \\
 $C_\rho$ &
     0.86740
 &
   0.66669
\\
   \hline \hline
 \end{tabular}
 \end{center}
 \caption{ The parameters that appear only inside the potential.}
  \label{beltana}
 \end{table}

\section{Partial wave amplitudes}
\label{partialwaveamplitudes}
To compare the $\pi N\rightarrow\pi N$ and $\gamma N\rightarrow\pi N$ amplitudes with experiment, we subject them to partial wave decompositions and then, in the case of $\pi N\rightarrow\pi N$, compute phase shifts.  This appendix contains formulae for the partial wave amplitudes and phase shifts.

\subsection{Pion-nucleon scattering}
\label{pionnucleonscatteringappendix}
To obtain the partial wave version of $\calT$, we first sandwich it with Dirac spinors and decompose it into isospin $1/2$ and $3/2$ components:
\be\bs
\calT_{ba}^{\bar{u}u}=(P_{1/2})_{ba}\calT_{1/2}^{\bar{u}u}+(P_{3/2})_{ba}\calT_{3/2}^{\bar{u}u}\eqn{decomposingintoiospincomponents}
\end{split}\ee
where $\calT_{ba}^{\bar{u}u}(k_f,p_f,k_i,p_i)=\bar{u}(\bp_f)\calT_{ba}(k_f,p_f,k_i,p_i)u(\bp_i)$.  When all the external particles are on-shell, $k_f^2=k_i^2=m_\pi^2$, $p_f^2=p_i^2=m_N^2$, and this allows the terms on the RHS of \eq{decomposingintoiospincomponents} to be expressed as,
\be\bs
\calT_t^{\bar{u}u}=\zeta_{1t}\bar{u}(\bp_f)u(\bp_i)+\zeta_{2t}\bar{u}(\bp_f)\gamma_0 u(\bp_i)\eqn{lahiffspreludetopartialwave}
\end{split}\ee
Calculating the partial wave amplitudes $T_{l'ljt}^{\bar{u}u}$ is then a matter of using $\zeta_1$, $\zeta_2$ in the following expression \cite{Lahiff:1999ur}:
\be\bs
\calT^{\bar{u}u}_{l'ljt}&=\frac{\e_\bark}{m_N}\pi\int_{-1}^{1}dx\left\{{\color{white}\frac{\frac{\frac{p}{}}{P}}{p}}\hspace{-0.4cm}P_l(x)\left[\zeta_{1t}(x)+\zeta_{2t}(x)\right]\right.\\&\,\,\,\,\,\,\left.+\left(\frac{\bark}{\e_\bark}\right)^2P_{l\pm 1}(x)\left[\zeta_{2t}(x)-\zeta_{1t}(x)\right]\right\}\delta_{l'l}\eqn{lyndhurst}
\end{split}\ee
where $x=\frac{\bp_i\cdot\bp_f}{|\bp_i||\bp_f|}$, $\e_\bark=\sqrt{\bark^2+m_N^2}+m_N$, and the on-shell relative momentum $\bark$ is given in \eq{therelativemomentumonshell}.  The $P$'s are Legendre polynomials and the $\pm$ sign refers to the total angular momentum $j=l\pm\half$.  

When a separable potential is used, putting the amplitude into the form of \eq{lahiffspreludetopartialwave} is very simple since it depends only on the total centre of mass momentum $p=(p_0,{\bf 0})$.  In that case $\zeta_{1t}$, $\zeta_{2t}$ are independent of $x$ and we obtain,  
\be\bs
&\calT^{\bar{u}u}_{00\half t}=\frac{2\pi\e_{\bark}}{m_N}\left(\zeta_{1t}+\zeta_{2t}\right)\\&
\calT^{\bar{u}u}_{11\half t}=\frac{2\pi\bark^2}{\e_{\bark}m_N}\left(\zeta_{2t}-\zeta_{1t}\right)
\end{split}\ee
The partial wave amplitudes are related to phase shifts and the inelasticity parameter $\eta$ by the formulae,
\be\bs
&\delta_{ljt}=\half\tan^{-1}\left(\frac{-\text{Re}(\alpha\calT^{\bar{u}u}_{l'ljt})}{\text{Im}(\alpha\calT^{\bar{u}u}_{l'ljt})+1}\right)\\&
\eta_{ljt}^2=[1+\text{Im}(\alpha\calT^{\bar{u}u}_{l'ljt})]^2+[\text{Re}(\alpha\calT^{\bar{u}u}_{l'ljt})]^2
\end{split}\ee
where
\be\bs
\alpha=\frac{m_N\bark}{8\pi^2p_0}
\end{split}\ee
Because $\calT_{l'ljt}^{\bar{u}u}$ has unitarity, $\eta$ must be equal to 1 for all $p_0>0$.  

\subsection{Pion photoproduction}
\label{pionphotoproductionappendix}
To compare the properly normalised $\gamma N\rightarrow\pi N$ amplitude $T_Z^{\mu a}$ with experiment, one first needs to sandwich it with Dirac spinors, isospin $\half$ states and contract it with a polarisation vector:
\be\bs
&\muui=\la \half m_t'|\bar{u}(\bp-\bk_f)T_Z^{\mu a}(k_f,p-k_f,p_i)\\&\,\,\,\,\,\,\,\,\times u(\bp_i)|\half m_t\ra\hat{\text{e}}_\mu(\lambda)
\end{split}\ee
where $m_t$ and $m_t'$ are the $z$-axis projections of the initial and final nucleon isospins.  The polarisation vector is given by,
\be\bs
&
\hat{\text{e}}_\mu(\pm1)=\mp\frac{1}{\sqrt{2}}(0,1,\pm i,0)
\end{split}\ee
where the argument $\lambda=\pm 1$ is the helicity; the amplitudes we calculate are independent of how this is chosen.  
When the kinematics are set according to \eq{howpeoplechoosethemomentumforphotproduction}, $\muui$ can be decomposed into products of ordinary numbers and rotationally invariant matrices as \cite{CGLN},
\be\bs
&-\frac{m_N}{4\pi p_0}\muuj= i\bsig\cdot\bhe(\lambda) J_{a m_t'm_t}^{(1)}\\&\,\,\,+\bsig\cdot\bhk_f\bsig\cdot [\bhq\times\bhe(\lambda)]J_{a m_t'm_t}^{(2)}\\&\,\,\,+i\bsig\cdot \bhq\bhk_f\cdot\bhe(\lambda)J_{a m_t'm_t}^{(3)}+i\bsig\cdot\bhk_f\bhk_f\cdot\bhe(\lambda)J_{a m_t'm_t}^{(4)}\eqn{nblsequationbfour}
\end{split}\ee
where $\bhk_f=\frac{\bk_f}{|\bk_f|}$ and $\bhq=\frac{\bq}{|\bq|}$.  In writing this it has been assumed that $\bar{u}$ and $u$ have the Bjorken and Drell \cite{Bjorken:Drell_1} normalisation.

Note that the LHS of \eq{nblsequationbfour} differs from NBL's \cite{Nozawa:1989pu} equation (B.4) by a factor of $-1$.  This discrepancy occurs because NBL's amplitude $M_{\pi N,\gamma N}^a$ consists of Feynman diagrams multiplied by $-i$.  This may be seen by deriving the Born terms in their equations (2.8b) and (2.8c) from Feynman rules.  Meanwhile, the amplitude $M_{a m_t'm_t}^{\bar{u}u}$ in the above \eq{nblsequationbfour} consists of Feynman diagrams multiplied by $i$ and some renormalisation constants.  To compensate, an extra minus sign has been included in this equation.  
 
 $J_{a m_t'm_t}^{(1)}$, $J_{a m_t'm_t}^{(2)}$, etcetera are used to calculate the invariant amplitudes,
\be\bs
\left(\begin{array}{c} E_{l^+}^{a m_t'm_t}\\ E_{l^-}^{a m_t'm_t}\\ M_{l^+}^{a m_t'm_t}\\ M_{l^-}^{a m_t'm_t}\end{array}\right)=\int_{-1}^1 dx D_l(x)\left(\begin{array}{c} J^{(1)}_{am_t'm_t} \\ J^{(2)}_{am_t'm_t} \\ J^{(3)}_{a m_t'm_t}\\ J^{(4)}_{a m_t'm_t}\end{array}\right)\eqn{thephysicalinvariantamplitudes}
\end{split}\ee
where $x=\cos\theta$ and $\theta$ is the same as that which appears in \eq{howpeoplechoosethemomentumforphotproduction}.  The matrix $D_l$ is given by,
\be\bs
&D_l(x)=\left(\begin{array}{cccc} a_lP_l & -a_lP_{l+1} & \frac{a_ll}{2l+1}Q_l & \frac{a_l(l+1)}{2l+3}Q_{l+1} \\ 
b_lP_l & -b_lP_{l-1}& -\frac{b_l(l+1)}{2l+1}Q_l & -\frac{b_ll}{2l-1}Q_{l-1}\\ c_lP_l& -c_lP_{l+1} & -\frac{c_l}{2l+1}Q_l & 0 \\ -d_lP_l & d_lP_{l-1}&\frac{d_l}{2l+1}Q_l & 0\end{array}\right)\eqn{thebigfuckinmatrix}
\end{split}\ee
where $a_l=\frac{1}{2(l+1)}$, $b_l=\frac{1}{2l}$, $c_l=\frac{1}{2(l+1)}$, $d_l=\frac{1}{2l}$, $Q_l=P_{l-1}-P_{l+1}$.  The $P$'s in \eq{thebigfuckinmatrix} and in $Q$ are Legendre polynomials and are functions of $x$.

Assuming the charged pion isospin matrices use the sign and normalisation convention $\tau_{\pm1}\equiv(\tau_1\pm i\tau_2)/\sqrt{2}$ and $\tau_0\equiv\tau_3$, the multipole amplitudes  are given by the following combinations of physical amplitudes \cite{Pfeil:1972pb}:
\be\bs
&M_{l^\pm}(0)=\frac{1}{2\sqrt{2}}\left[M_{l^\pm}(\gamma n\rightarrow \pi^- p)+M_{l^\pm}(\gamma p\rightarrow \pi^+ n)\right]\\&
M_{l^\pm}(1/2)=M_{l^\pm}(\gamma p\rightarrow \pi^0 p)\\&\,\,\,\,\,\,\,\,\,-\frac{1}{2\sqrt{2}}\left[3M_{l^\pm}(\gamma n\rightarrow \pi^- p)-M_{l^\pm}(\gamma p\rightarrow \pi^+ n)\right]\\&
\eqn{relatingthephysicalmultiploestothesupidones}
\end{split}\ee
where $M_{l^\pm}(\gamma n\rightarrow \pi^- p)$ is equal to $M_{l^\pm}^{am_t'm_t}$ with $a=-1$, $m_t'=\half$, $m_t=-\half$.  Similarly, $M_{l^\pm}(\gamma p\rightarrow\pi^+ n)$ is equal to $M_{l^\pm}^{am_t'm_t}$ with $a=1$, $m_t'=-\half$, $m_t=\half$ and $M_{l^\pm}(\gamma p\rightarrow \pi^0 p)$ is equal to $M_{l^\pm}^{am_t'm_t}$ with $a=0$, $m_t'=\half$, $m_t=\half$.  
It is also conventional to decompose the $t=1/2$ amplitudes into ``proton" and ``neutron" pieces:
\be\bs
&_pM_{l^\pm}(1/2)=M_{l^\pm}(0)+\frac{1}{3}M_{l^\pm}(1/2)\\&_nM_{l^\pm}(1/2)=M_{l^\pm}(0)-\frac{1}{3}M_{l^\pm}(1/2)
\eqn{drechselandtiatorsstuff}
\end{split}\ee
Relations identical to those appearing in \eq{relatingthephysicalmultiploestothesupidones} and \eq{drechselandtiatorsstuff} also apply to the $E$ amplitudes.  To check for mistakes in the computer program used to calculate the multipole amplitudes, we verified that it correctly reproduces the Born term results of Laget \cite{Laget:1981jq}.

\section{The coupling constant}\label{fixingthecouplingconstant}
To evaluate the coupling constant of the dressed $\pi NN$ vertices, we compare the pole (second) term from the first of equations (\ref{extendedtoroperparticles}) to a similar diagram that has bare vertices with $g_{\pi NN}$ substituted for $g_{\pi NN}^{(0)}$  and a bare propagator for a physical mass nucleon.   The dressed coupling constant may be extracted by equating the residues of the two diagrams at the nucleon pole.

The properly normalised $\pi N\rightarrow\pi N$ pole term is,
\be\bs
\calT_\text{Pole}(k_f,p_f,k_i,p_i)&=iZ_2{\bff}(k_f,p_f,p)\bg(p)\bar{\bff}(k_i,p,p_i)
\eqn{thedressedpolediagram}
\end{split}\ee
where $p=p_i+k_i=p_f+k_f=(p_0,{\bf 0})$.
Putting the external particles on shell (that is, choosing $p_i^2=p_f^2=m_N^2$, $k_i^2=k_f^2=m_\pi^2$) and sandwiching $\calT_\text{Pole}$ with Dirac spinors allows it to be expressed in terms of the $\Lambda^\pm$ operators, giving
\be\bs
\calT_\text{Pole}^{\bar{u}u} &=\calT_\text{Pole}^+ U^++\calT_\text{Pole}^-U^-
\end{split}\ee
where $\calT_\text{Pole}^{\bar{u}u}=\bar{u}(\bp_f)\calT_\text{Pole}u(\bp_i)$ and 
\be\bs
&T_\text{Pole}^\pm(p_0)=-I_\tau Z_2\sum_{\beta\alpha}f_{\beta}^\mp(p_0) g_{\beta\alpha}^\mp(p_0) \barf_\alpha^\mp(p_0)\\&
U^\pm=\bar{u}(\bp_f)\Lambda^\pm u(\bp_i)\eqn{goingtothepoleoftpole}
\end{split}\ee
The $f_\alpha^\pm(p_0)$ factors are ordinary numbers (not matrices) given by expressions that are straightforward to derive, while $I_\tau$ is an isospin factor.  Now, the negative energy propagator $\bg^-$ doesn't have poles at the particle masses and so the residue of $\calT_\text{Pole}$ at $p_0=m_N$ is,
\be\bs
&\mathop{\text{Res}}_{p_0=m_N}\calT_\text{Pole}^{\bar{u}u}=-I_\tau Z_2\sum_{\alpha\beta}{f_{\alpha}^{+}}(m_N)f_\beta^+(m_N)\sqrt{Z_\alpha Z_\beta}U^-\eqn{theresidueofthecalculateddiagram}
\end{split}\ee
The diagram to which this should be compared is,
\be\bs
&V_\text{Pole}^{\bar{u}u}=-I_\tau g_{\pi NN}^2\bar{u}(\bp_f)\gamma_5\frac{\psl+m_N}{p^2-m_N^2+i\e}\gamma_5 u(\bp_i)
\end{split}\ee
When all particles are on shell and the diagram is multiplied by Dirac spinors, pseudovector coupling is equivalent to pseudoscalar.  Using pseudoscalar coupling therefore causes no loss of generality and in either case the residue at $p_0=m_N$ is,
\be\bs
\mathop{\text{Res}}_{P_0=m_N} V_\text{Pole}^{\bar{u}u}=-I_\tau g_{\pi NN}^2U^-
\end{split}\ee
Equating this to the RHS of \eq{theresidueofthecalculateddiagram}, we see that,
\be\bs
g_{\pi NN}&=\sqrt{Z_2}f_N^+(m_N)\sqrt{Z_N}+\sqrt{Z_2}f_R^+(m_N)\sqrt{Z_R}
\end{split}\ee
The bare coupling constant $g_{\pi NN}^{(0)}$ should be set so that $g_{\pi NN}\approx 13.02$.

\end{appendix}


\bibliographystyle{apsrev4-1.bst}
\bibliography{ref}

\end{document}